%% file: Semiparametric_Bayesian_inference_for_causal_mediation_in_cluster_randomized_trials.tex
\documentclass[12pt]{article}
\usepackage{woojung}
\setlist{leftmargin=0.1in}

\title{Semiparametric Bayesian inference for causal mediation in cluster randomized trials}

\author[1]{Woojung Bae}
\author[1]{Michael J. Daniels \footnote{Corresponding author}}
\author[2]{Joseph W. Hogan}
\author[3,4]{Rajesh Vedanthan}
\author[2]{Stavroula Chrysanthopoulou}

\affil[1]{Department of Statistics, University of Florida}
\affil[2]{Department of Biostatistics, Brown University}
\affil[3]{Institute for Excellence in Health Equity, NYU Grossman School of Medicine}
\affil[4]{Department of Population Health, Department of Medicine, NYU Grossman School of Medicine}

\myexternaldocument{Supp - Semiparametric Bayesian inference for causal mediation in cluster randomized trials}

\begin{document}
\maketitle

\setlength{\parskip}{2pt}
\setlength{\abovedisplayskip}{2pt}
\setlength{\belowdisplayskip}{2pt}
\setlength{\abovedisplayshortskip}{2pt}
\setlength{\belowdisplayshortskip}{2pt}

\begin{abstract}
    Cluster randomized trials (CRTs) are frequently used to evaluate interventions, yet conducting causal mediation analysis in these settings remains challenging, particularly when the mediator is measured at the cluster level and the number of clusters is small. Standard inference methods often rely on asymptotic assumptions that fail in finite-sample settings, leading to biased variance estimation and invalid confidence intervals. In this paper, we propose a robust inference framework for causal mediation analysis in CRTs. We utilize parametric Bayesian models for the outcome and mediator to ensure computational efficiency and interpretability. Crucially, to quantify uncertainty, we specify a novel similarity-weighted Bayesian bootstrap (SWBB) with a ‘distance’ metric between clusters; this avoids the need for restrictive parametric assumptions and allows the model to borrow more information from ‘closer’ clusters. By combining observed data models with causal assumptions, our approach accurately estimates natural direct and indirect effects even with limited clusters. Simulation studies demonstrate that our method achieves nominal coverage probability across diverse scenarios. We illustrate the practical utility of our approach by assessing mediation in a CRT in Kenya.
\end{abstract}

\clearpage
\section{Introduction} \label{SWBB.introduction}
    Cluster randomized trials (CRTs) allocate treatments to groups of subjects rather than to individuals. These trials are commonly used to evaluate complex interventions in health services research \citet{hayes2017cluster}. This design is particularly appropriate when interventions are implemented at the level of organizations or regions, such as during hospital restructuring, guideline implementation, or the introduction of new models of care.
    
    \input{figure/figureSWBB_DAG}
    The causal graph in Figure~\ref{SWBB.causalstructure} visualizes the structure of a Cluster Randomized Trial (CRT) where the treatment $Z_{j}$ and mediator $M_{j}$ are defined at the cluster level, while the outcome $Y_{ji}$ is measured at the individual level. A bi-directional arrow connects the cluster-level confounders $\boldsymbol{V}_{j}$ and individual-level confounders $\boldsymbol{C}_{ji}$, indicating that these confounding factors may be correlated with each other. The solid arrows represent the specific causal pathways of interest: path $(a)$ denotes the natural direct effect of the intervention $Z_{j}$ on the outcome $Y_{ji}$; path $(b)$ captures the effect of the intervention on the mediator $M_{j}$; and path $(c)$ represents the effect of the mediator on the outcome. The total effect of exposure $Z_{j}$ is the aggregate of these direct and indirect pathways. Dotted lines originating from the grouped confounders indicate the confounding effects that need to be adjusted for. 

    To incorporate covariates and improve the efficiency of estimating treatment effects in cluster randomized trials (CRTs), researchers have utilized various approaches, including generalized estimating equations (GEE), covariate-adjusted residuals estimation (CARE), targeted maximum likelihood estimation (TMLE), and augmented GEE (Aug-GEE) \citep{liang1986longitudinal, gail1996design, vander2011targeted, stephens2012Augmented, balzer2016adaptive, crespi2016improved, balzer2019new, su2021model, benitez2023defining, son2024spatial, hong2025multivariate}. Recently, \citet{cheng2024semiparametric} expanded on this by developing a model-robust frequentist framework for covariate adjustment, utilizing efficient influence functions and machine learning to ensure valid inference on both cluster-average and individual-average treatment effects even under working model misspecification. Similarly, \citet{benitez2023defining} focused on defining and estimating effects in CRTs using Pearl's nonparametric structural equation model \citep{pearl2009causality}. While \citet{benitez2023defining} addressed the selection of appropriate weights for weighted sums of outcomes—defining effects by differentiating these weights—their work, like others in this category, primarily targets the total treatment effect. It does not cover mediation analysis, and their definition of cluster-level or individual-level effects differs from our approach, where we define effects as the expected value of the counterfactual outcome given specific confounders while marginalizing out others.

    When the focus shifts to mechanisms of action involving individual-level mediators, several frequentist and Bayesian approaches have been proposed. \citet{vanderweele2008ignorability} and \citet{vanderweele2010direct} employed the potential outcomes framework to establish causal assumptions for identifying direct and indirect effects in CRTs and clustered longitudinal data. Building on this, \citet{kelcey2017statistical} developed closed-form expressions for the variance and power of detecting causally defined indirect effects in two-level studies with individual-level mediators. To address heterogeneity, \citet{park2015bayesian} demonstrated that Bayesian approaches could account for heterogeneous treatment or mediator effects without bias, while \citet{reardon2013what} utilized instrumental variable (IV) methods to examine multiple proposed mediators. More recently, advanced semiparametric and nonparametric frameworks have emerged to handle complex data structures such as interference and spillover effects. \citet{wang2024model} developed a semiparametric efficiency theory to identify natural indirect, individual, and spillover mediation effects using a single individual-level mediator. Furthermore, \citet{ohnishi2025bayesian} proposed a Bayesian nonparametric (BNP) framework using the Nested Dependent Dirichlet Process Mixture (nDDPM) prior to decompose total causal effects in the presence of interference and multiple individual-level mediators.
    
    Despite the progress in individual-level mediation, research focusing specifically on cluster-level mediators—or the distinction between cluster- and individual-level mechanisms—remains less developed. \citet{pituch2012distinguishing} highlighted the critical theoretical divergence in CRTs, arguing that the nature of the mediating variable determines whether a treatment effect can be mediated by an individual-level or exclusively a cluster-level mediator. While general Bayesian nonparametric (BNP) methods for mediation have been introduced—such as Bayesian causal mediation forests \citep{linero2022mediation}, varying coefficient models \citep{ting2023estimating}, and Dirichlet process mixture variations \citep{daniels2012bayesian, kim2017framework, kim2019bayesian, bae2024bayesian, roy2024bayesian, ohnishi2025bayesian}—most methods are not directly adapted to the specific hierarchical structure of CRTs. Consequently, few approaches effectively address cluster-specific mediation effects or provide a unified framework that simultaneously accommodates cluster-level mediators alongside individual-level outcomes. This highlights a gap in the literature regarding the estimation of cluster-specific effects, which we define as the expected value of the counterfactual outcome given cluster-level confounders, independent of individual-level variations.

    Various modifications to the Bayesian bootstrap (BB) have been proposed in the literature to address specific structural or computational challenges. For instance, \citet{makela2018bayesian} developed a two-staged BB specifically for CRTs, in which clusters are sampled first, followed by individuals within each cluster. Their primary focus was to account for population clusters or strata that exist but are never sampled. This differs from our setting, where cluster or stratum information is fixed and known, and our goal is to partially pool information across them. Other approaches, such as the `bag-of-little bootstraps' proposed by \citet{kleiner2014scalable} and \citet{barrientos2020bayesian}, have aimed to scale the bootstrap to large datasets. These methods approximate the overall bootstrap distribution by running separate bootstraps on sub-samples and combining the results. However, our interest lies specifically in estimating the confounder distribution rather than the overall data distribution. In related work, \citet{oganisian2022hierarchical} introduced a hierarchical Bayesian bootstrap (HBB) prior to estimate cluster/stratum-specific confounder distributions. Their approach, based on the hierarchical Dirichlet Process (HDP), allows for borrowing of confounder information across clusters/strata in a principled manner. However, their method is somewhat restrictive in terms of not being able to explicitly favor borrowing information from the `closer' clusters. Additionally, several `smoothed' bootstraps have been developed, such as those proposed in \citet{efron1983leisurely, silverman1987bootstrap, wang1995optimizing, oganisian2022hierarchical}. These smoothed bootstraps employ a point-mass empirical distribution with a parametric kernel to induce smoothness in the resampled data.
    
    We extend previous work on mediation analysis in CRTs and introduce semiparametric Bayesian models. For the latter, we introduce a similarity-weighted BB (SWBB) that incorporates a `distance' metric between clusters. Our approach enables the estimation of joint distributions of cluster-level confounders and individual-level confounders while leveraging more information from the clusters that are `closer'. Our BB approach offers several advantages: (1) it maintains the flexibility of the BB by not assuming any specific parametric form for the confounder distributions, (2) it is fully conjugate, meaning it can handle various outcome and mediator modeling approaches without restrictions or assumptions, and (3) it is agnostic to the choice of outcome model, making it compatible with several popular approaches mentioned earlier. We employ parametric models for the outcome and mediators to facilitate clear interpretation and inference. This enables us to appropriately account for dependencies within clusters by explicitly modeling the correlation structure, while allowing for the straightforward incorporation of various covariates and adjustment for confounding variables.

    We demonstrate the utility of our approach using data from the Bridging Income Generation with Group Integrated Care (BIGPIC) trial, a cluster-randomized trial in Western Kenya aimed at reducing cardiovascular risk \citep{vedanthan2017bridging, vedanthan2021group}. In this study, $J=24$ health facility sites were randomized to four arms: (1) usual care, (2) microfinance (MF) only, (3) group medical visits (GMV) only, and (4) an integrated GMV-MF model. The primary outcome is the one-year change in systolic blood pressure (SBP). We investigate how the transition to integrated care affects SBP through changes in the structural characteristics of participants' social networks. Specifically, we conduct pairwise comparisons (GMV given MF and vice versa) to isolate the additive effects of the integrated model. We evaluate four sociometric network characteristics measured at 12 months—density, transitivity, cohesion, and average path length—as potential mediators. To address the challenge of limited cluster-level data, we utilize a single-mediator framework for each characteristic. This allows us to maintain the robustness of the SWBB approach and leverage its cross-cluster borrowing strategy to achieve precise causal estimates without the risk of over-parameterization associated with multiple-mediator models in small samples.

    The article is structured as follows. Section~\ref{SWBB.causal} defines the potential outcomes and causal estimands—Natural Indirect Effects (NIE) and Natural Direct Effects (NDE)—at the individual, cluster, and population levels, and introduces the sequential ignorability (SI) assumptions necessary for identification in a cluster-randomized trial (CRT) framework. Section~\ref{SWBB.similarity} presents the core methodological contribution: the SWBB framework for modeling the joint distribution of cluster- and individual-level confounders. This section details the distance-based, adaptive information-borrowing mechanism designed to improve estimation in the presence of cluster-level heterogeneity and data sparsity. Section~\ref{SWBB.outcome} describes the modeling strategy for the observed data, specifying the parametric mixed-effects and fixed-effects regressions used to characterize the individual-level outcome and cluster-level mediator, respectively. Section~\ref{SWBB.posterior} outlines the posterior computation procedure, which integrates the outcome and mediator models over the SW-BB confounder distribution to obtain causal mediation estimands at the cluster, individual, and population levels. Section~\ref{SWBB.similarity} presents the core methodological contribution: a SWBB framework for modeling the joint distribution of cluster- and individual-level confounders. This section details the adaptive information-borrowing mechanism that utilizes a distance-based metric to improve estimation in the presence of cluster-level heterogeneity. Section~\ref{SWBB.outcome} specifies the parametric mixed-effects and fixed-effects models used to characterize the individual-level outcome and cluster-level mediator, respectively. Section~\ref{SWBB.posterior} specifies the parametric mixed-effects and fixed-effects models used to characterize the individual-level outcome and cluster-level mediator, respectively. In Section~\ref{SWBB.simulation}, we conduct simulation studies across various cross-level dependence scenarios to evaluate the bias, precision, and coverage of the SWBB relative to existing Bayesian bootstrap specifications. The proposed approach is then applied to the BIGPIC trial data in Section~\ref{SWBB.BIGPIC}, where we investigate how sociometric social network characteristics mediate the additive effects of integrated medical and financial interventions on blood pressure. Finally, Section~\ref{SWBB.discussion} discusses the findings, the inferential implications for small-cluster designs, and potential avenues for future research.
    
\section{Causal estimand and identification assumptions} \label{SWBB.causal} 
    In CRTs, treatments are assigned to entire facilities or clusters of subjects, rather than individual subjects. The treatment is often assumed to influence the individual-level outcome through a mediating variable measured at the cluster level. Furthermore, we observe two sets of baseline confounders for causal mediation, one at the individual-level and the other at the cluster-level. In CRTs, we often are interested in estimating mediation effects at the individual-level and/or at the cluster-level.

\subsection{Potential outcomes and causal estimand} \label{SWBB.causal.potential}
    Consider a study design where we observe an outcome $Y_{ji}$ at the individual level and a mediator $M_{j}$ at the cluster level, with each cluster assigned to one of two treatments $Z_{j} \in \left\{ 0, 1 \right\}$. Let $\bbX_{ji} = \left( \bbV_{j}, \bbC_{ji} \right)$ denote the vector of pre-treatment confounders, where $\bbC_{ji} \in \IR^{p_{\bbC}}$ represents individual-level confounders and $\bbV_{j} \in \IR^{p_{\bbV}}$ represents cluster-level confounders. We assume that $\bbV_{j}$ varies across clusters, yielding $J$ distinct observed vectors. We adopt the potential outcome framework \citep{rubin1974estimating} to define causal effects. Let $M_{j} \left( z \right)$ denote the potential value of the mediator if cluster $j$ is assigned to treatment $z$. The observed mediator is given by $M_{j} = Z_{j} M_{j} (1) + \left( 1 - Z_{j} \right) M_{j} (0)$. Similarly, let $Y_{ji} \left( z, M_{j} \left( z' \right) \right)$ denote the potential outcome for individual $i$ in cluster $j$ under treatment $z$ and mediator value $M_{j}(z')$. The observed outcome corresponds to the potential outcome under the assigned treatment, expressed as $Y_{ji} = Z_{j} Y_{ji} ( 1, M_{j} (1) ) + \left( 1 - Z_{j} \right) Y_{ji} ( 0, M_{j} (0) )$. 
    
    We define our causal estimates of interest—natural direct and indirect effects—at the individual-level, cluster-level, and population-level. Unlike the conditional nature of the individual and cluster estimates, the population-level effects represent the aggregate impact, obtained by marginalizing over the distribution of individual- and cluster-level confounders. We formalize these estimands as follows. Conditional on individual-level confounders $\bbC_{ji} = \bbc$, the individual-specific effects are defined as:
    \begin{align*}
        \nnie \left( \bbc \right) &= \mE[Y_{ji} ( 1, M_{j} (1) ) - Y_{ji} ( 1, M_{j} (0) ) \mid \bbC_{ji} = \bbc], \\
        \nnde \left( \bbc \right) &= \mE[Y_{ji} ( 1, M_{j} (0) ) - Y_{ji} ( 0, M_{j} (0) ) \mid \bbC_{ji} = \bbc].
    \end{align*}
    Here, the natural indirect effect (NIE) quantifies the treatment effect transmitted through the mediator, while the natural direct effect (NDE) captures the effect independent of the mediator conditional on individual-level confounders $\bbC_{ji} = \bbc$. Similarly, conditioning on cluster-level confounders $\bbV_{j} = \bbv$ yields the cluster-specific estimates $\nnie \left( \bbv \right) $ and $\nnde \left( \bbv \right) $. Finally, the population-level effects are defined by marginalizing over these confounders:
    \begin{align*}
        \nnie &= \mE[Y_{ji} ( 1, M_{j} (1) ) - Y_{ji} ( 1, M_{j} (0) )], \\
        \nnde &= \mE[Y_{ji} ( 1, M_{j} (0) ) - Y_{ji} ( 0, M_{j} (0) )].
    \end{align*}
    Throughout the remainder of the article, we suppress the subject index $i$ and cluster index $j$ for notational simplicity and focus on assessing the natural effects \citep{robins1992identifiability, pearl2013direct}. Specifically, we denote the cluster-level variables as $M$ (mediator), $Z$ (treatment), and $\bbV$ (confounders), and the individual-level variables as $Y$ (outcome) and $\bbC$ (confounders).

\subsection{Identifying assumptions and inference on causal effects} \label{SWBB.causal.identifying}
    To identify the proposed causal estimands from the observed data, we rely on two primary structural assumptions. First, we invoke a Cluster-level Stable Unit Treatment Value Assumption (SUTVA), adapted from \citet{ohnishi2025bayesian} for a cluster-level mediator. Letting $\mathbf{Z}$ and $\mathbf{M}$ denote the vectors of treatment assignments and mediators across all $J$ clusters, we assume \eqref{SWBB.causal.eq.1} no cross-cluster interference—meaning a cluster's potential mediator and its individuals' potential outcomes depend solely on the local cluster assignment—and \eqref{SWBB.causal.eq.2} consistency between potential and observed variables:
    \begin{gather}
        M_{j}(\mathbf{Z}) = M_{j}(Z_{j}) \quad \text{and} \quad Y_{ji}(\mathbf{Z}, \mathbf{M}) = Y_{ji}(Z_{j}, M_{j}), \label{SWBB.causal.eq.1} \tag{A1.1} \\
        M_{j} = M_{j}(z) \quad \text{and} \quad Y_{ji} = Y_{ji}(z, m) \quad \text{if } Z_{j} = z \text{ and } M_{j} = m. \label{SWBB.causal.eq.2} \tag{A1.2}
    \end{gather}
    
    Second, we invoke the sequential ignorability (SI) assumption \citep{imai2010identification} to rule out unmeasured confounding. Let $\bbC$ and $\bbV$ denote the sets of baseline individual- and cluster-level confounders, respectively. In the absence of exposure-induced confounding, we assume:
    \begin{gather}
        \left\{ Y (z', m) , M (z) \right\} \indep Z \mid \bbC = \bbc, \bbV = \bbv, \label{SWBB.causal.eq.3} \tag{A2.1} \\
        Y (z', m) \indep M (z) \mid Z = z, \bbC = \bbc, \bbV = \bbv, \label{SWBB.causal.eq.4} \tag{A2.2}
    \end{gather}
    where $z, z' \in \left\{ 0,1 \right\}$. In a cluster-randomized design, condition \eqref{SWBB.causal.eq.3} is inherently satisfied by the treatment assignment mechanism. We further assume standard positivity ($\mP[Z = z \mid \bbC = \bbc, \bbV = \bbv ] > 0$ and $\mP[M = m \mid Z = z, \bbC = \bbc, \bbV = \bbv ] > 0$) for all combinations of $z, z', \bbc, \bbv$, and $m$ over their respective supports. 
    
    Under these assumptions, the nested potential outcome is identified (see Section~\ref{SWBB.supp.identification} of the Supplementary Materials) via the mediation formula:
    {
    \fontsize{11}{11}\selectfont
    \begin{align*}
        \mE[ Y ( z, M (z') ) \mid \bbC = \bbc, \bbV = \bbv]
        & = \int \mE[ Y \mid M = m', Z = z, \bbC = \bbc, \bbV = \bbv] \ndF_{M \mid Z = z', \bbC = \bbc, \bbV = \bbv} \left( m' \right).
    \end{align*}
    }
    
    Following this identification result, we obtain point estimates for causal effects at various levels of aggregation by marginalizing over the empirical distributions of the relevant confounders. The specific level of the estimated mediation effect—whether at the individual, cluster, or population level—is determined by the subset of confounders being marginalized:
    \begin{itemize}
        \item \textit{Cluster-level effects;} Identified by marginalizing over the individual-level confounders $\bbC$ for a given cluster profile $\bbV = \bbv$:
        {
        \fontsize{10}{11}\selectfont
        \begin{align*}
            \mE[Y ( z, M (z') ) \mid \bbV = \bbv]
            =
            \iint \mE[ Y \mid M = m', Z = z, \bbC = \bbc, \bbV = \bbv] \ndF_{M \mid Z = z', \bbC = \bbc, \bbV = \bbv} \left( m' \right) \ndF_{\bbC \mid \bbV = \bbv} \left( \bbc \right).
        \end{align*}
        }
        
        \item \textit{Individual-level effects;} Identified by marginalizing over the cluster-level confounders $\bbV$ for a given individual profile $\bbC = \bbc$:
        {
        \fontsize{10}{11}\selectfont
        \begin{align*}
            \mE[Y ( z, M (z') ) \mid \bbC = \bbc]
            =
            \iint \mE[ Y \mid M = m', Z = z, \bbC = \bbc, \bbV = \bbv] \ndF_{M \mid Z = z', \bbC = \bbc, \bbV = \bbv} \left( m' \right) \ndF_{\bbV \mid \bbC = \bbc} \left( \bbv \right).
        \end{align*}
        }

        \item \textit{Population-level effects;} Identified by marginalizing over the joint distribution of $\{\bbC, \bbV\}$:
        {
        \fontsize{10}{11}\selectfont
        \begin{align*}
            \mE[Y ( z, M (z') )]
            & =
            \iiint \mE[ Y \mid M = m', Z = z, \bbC = \bbc, \bbV = \bbv]  \ndF_{M \mid Z = z', \bbC = \bbc, \bbV = \bbv} \left( m' \right) \ndF_{\bbC \mid \bbV = \bbv} \left( \bbc \right) \ndF_{\bbV} \left( \bbv \right).
        \end{align*}
        }
        
    \end{itemize}

\section{The similarity-weighted Bayesian bootstrap} \label{SWBB.similarity}
    To estimate cluster-level, individual-level, and population-level mediation effects, we need to estimate the joint distribution of cluster-level confounders and individual-level confounders, $\mP[\bbC = \bbc,  \bbV = \bbv ]$. From a causal perspective, the confounder distributions are just a nuisance parameter, but we want to model it as flexibly as possible. The specifications for the outcome and mediator models are detailed in Section~\ref{SWBB.outcome}.

    Previous research on Bayesian causal inference for marginal causal estimation has utilized Rubin's BB approach, where the confounder distribution is modeled as a point mass at each observed covariate value with unknown weights. Improper Dirichlet priors are placed over the simplex \citep{rubin1981bayesian, kleiner2014scalable, makela2018bayesian, barrientos2020bayesian, oganisian2022hierarchical}. These approaches offer several advantages, including flexibility, uncertainty quantification, and computational simplicity. However, the BB approach encounters challenges when dealing with sparse clusters, where only a few unique values of $\bbC$ (the confounders) are observed. To address this issue, \citet{oganisian2022hierarchical} introduced the HBB, which allows for partial pooling of estimates for $\mP[\bbC = \bbc \mid \bbV = \bbv]$ when $V$ is a cluster level covariate taking on values $\bbv \in \left\{ 1, \ldots, J \right\}$. This extends the BB approach while preserving its desirable properties. However, HBB has a limitation in that it does not explicitly allow borrowing of information from `closer' (to be defined) clusters. Borrowing information from different clusters with equal weights can be problematic because clusters may share certain characteristics with some but possess distinct characteristics compared to others. We propose an alternative approach, the SWBB. SWBB enables us to borrow more information from `similar' clusters, where similarity is quantified by calculating the `distance' between clusters. With this information, we define cluster-level, individual-level, and population-level mediation effects. We provide a detailed description of our approach next.
    
    Let $\Sv = \bigcup_{l=1}^{J} \Sv[l]$ denote the set of observed cluster-level confounders where $\Sv[l] = \{ \bbV_{l} \}$ is a singleton set, and let $\Sc[j] = \bigcup_{i=1}^{n_{j}} \Sc[ji]$ be the set of individual-level confounders within cluster $j$ where $\Sc[ji] = \{ \bbC_{ji} \}$. To formally identify the joint distribution $\mP[\bbC = \bbc_{ji}, \bbV = \bbv_{l}]$, we assume that the specific individual confounder value is independent of the cluster-level profile, given its cluster membership: $\{ \bbc_{ji} \indep \bbv_{l} \mid \bbc \in \Sc[j] \}$. Under this assumption, the joint probability mass function (PMF) decomposes as:
    \begin{align}
        \mP[\bbC = \bbc_{ji}, \bbV = \bbv_{l}]
        &= \mP[\bbV = \bbv_{l}]
        \mP[\bbC \in \Sc[j] \mid \bbV = \bbv_{l}]
        \mP[\bbC = \bbc_{ji} \mid \bbC \in \Sc[j]].
        \label{SWBB.swbb.eq.1}
    \end{align}
    
    This decomposition is formally justified by the following steps, noting that $\Sc[ji] \subset \Sc[j]$:
    \begin{align*}
        \mP[\bbc \in \Sc[ji], \bbv \in \Sv[l] ]&= \mP[\bbc \in { \Sc[ji] \cap \Sc[j] }, \bbv \in \Sv[l] ] \nonumber \\
        &= \mP[\bbc \in \Sc[ji], \bbc \in \Sc[j], \bbv \in \Sv[l] ] \nonumber \\
        &= \mP[\bbc \in \Sc[ji] \mid \bbc \in \Sc[j], \bbv \in \Sv[l] ] \mP[\bbc \in \Sc[j] \mid \bbv \in \Sv[l] ] \mP[\bbv \in \Sv[l] ] \nonumber \\
        &= \mP[\bbc \in \Sc[ji] \mid \bbc \in \Sc[j] ] \mP[\bbc \in \Sc[j] \mid \bbv \in \Sv[l] ] \mP[\bbv \in \Sv[l] ].
    \end{align*}
    
    For the cluster-level confounder $\Sv$, we specify a BB $\mP[\bbv][\Sv] = \sum_{l=1}^{J} \rho_{l} \delta_{\bbV_{l}} \left( \bbv \right)$ where $\delta_{x} ( \cdot )$ denotes the degenerate distribution at $x$. For compactness we sometimes denote these as simply $\nP_{\Sv}$ and $\delta_{x}$. Here, $\bbrho = \left( \rho_{1}, \ldots, \rho_{J} \right)$ are considered unknown parameters that completely determine $\nP_{\Sv}$ and the weight vector lives in the simplex, $\bbrho \in \left\{ \IR^{J} : \rho_{l} > 0 \; \forall l = 1, \ldots, J, \; \& \; \sum_{l=1}^{J} \rho_{l} = 1 \right\}$, the BB places an improper Dirichlet prior over this space $\bbrho \sim \mdir{\bbzero_{J}} \approx \prod_{l=1}^{J} \rho_{l}^{-1}$. This is a conjugate model with posterior $\bbrho \mid \bbV \sim \mdir{\bbone_{J}}$ where $\bbone_{J}$ is the $J$-dimensional vector of ones:
    \begin{align} \label{SWBB.swbb.eq.2}
        \mP[\bbrho \mid \Sv ]
        \propto 
        \left\{ \prod_{l=1}^{J} \mP[\bbv \in \Sv[l] ; \bbrho ] \right\} \mP[\bbrho ] 
        \propto 
        \left\{ \prod_{l=1}^{J} \rho_{l} \right\} \mP[\bbrho ] 
        \propto 
        \left\{ \prod_{l=1}^{J} \rho_{l} \right\} \left\{ \prod_{l=1}^{J} \rho_{l}^{- 1} \right\}
        \propto 
        \prod_{l=1}^{J} \rho_{l}^{1 - 1}.
    \end{align}
    This is the kernel of $\mdir{\bbone_{J}}$ and is discrete with an atom at each of the observed $\Sv$. 
    
    We model the individual-level confounder within the cluster-level $j$, $\Sc[ji]$ given $\Sc[j]$ again using a BB, $\mP[\bbc][\Sc[j]] = \sum_{i=1}^{n_{j}} \pi_{i \mid \left( j \right)} \delta_{\bbC_{ji}} \left( \bbc \right)$ where $\bbpi_{\left( j \right)} = \left( \pi_{1 \mid \left( j \right)}, \ldots, \pi_{n_{j} \mid \left( j \right)} \right) \in \left\{ \IR^{n_{j}} : \pi_{i \mid \left( j \right)} > 0 \; \forall i = 1, \ldots, n_{j}, \; \& \; \sum_{j=1}^{n_{j}} \pi_{i \mid \left( j \right)} = 1 \right\}$ for each $j = 1, \ldots, J$. With an improper Dirichlet prior over this space $\bbpi_{\left( j \right)} \sim \mdir{\bbzero_{n_{j}}} \approx \prod_{i=1}^{n_{j}} \left( \pi_{i \mid \left( j \right)} \right)^{-1}$ for each $j = 1, \ldots, J$, the posterior of $\bbpi_{\left( j \right)}$ can be drawn from $\bbpi_{\left( j \right)} \mid \Sc[j] \sim \mdir{\bbone_{n_{j}}}$ and it can be derived in a similar manner to equation \ref{SWBB.swbb.eq.2}.

    For $\mP[\bbc \in \Sc[j] \mid \bbv \in \Sv[l] ]$, it is important to note that we only observe the data when $j = l$, and consequently, it is not identifiable from the observed data when $j \ne l$. We address this by incorporating a prior that allows for borrowing more confounder information from clusters that are considered `closer'. We quantify the closeness by computing the `distance' between clusters and specify a prior for $\mP[ \bbc \in \Sc[j] \mid \bbv \in \Sv[l] ]$ that is a function of the `distance'. We define the distance between the cluster $l$ and the cluster $j$ by 
    \begin{align*}
        \ndist_{lj} = \exp \left\{ - \frac{1}{\chi} \left( \zeta \xi_{lj}^{\bbV} + \left( 1 - \zeta \right) \xi_{lj}^{\bbC} \right) \right\}
    \end{align*}
    where $\chi > 0$ is a constant to scale the desired rate of decay and $\zeta \in \left[ 0, 1 \right]$. Here, $\xi_{lj}^{\bbV}$ and $\xi_{lj}^{\bbC}$ represent the distances between cluster $l$ and cluster $j$ for cluster-level and individual-level confounders, respectively, defined as:
    \begin{gather*}
        \xi_{lj}^{\bbV} = \frac{1}{p_{\bbV}} \sum_{q=1}^{p_{\bbV}} \norm{V_{lq} - V_{jq}}
        \; \text{and} \;
        \xi_{lj}^{\bbC} = \frac{1}{p_{\bbC} n_{l} n_{j}} \sum_{h=1}^{n_{l}} \sum_{i=1}^{n_{j}} \sum_{q=1}^{p_{\bbC}} \norm{C_{lhq} - C_{jiq}},
    \end{gather*}
    where $p_{\bbV}$ is the dimension of the cluster-level confounder vector, $p_{\bbC}$ is the dimension of the individual-level confounder vector and $\norm{\cdot}$ denotes a standard norm (e.g., the $L_1$ or $L_2$ norm).    
    
    Importantly, while $\xi_{lj}^{\bbC}$ measures the distance between two distinct clusters when $l \neq j$, it provides a natural measure of within-cluster heterogeneity when $l=j$. In this case, $\xi_{ll}^{\bbC}$ represents the average dissimilarity between all pairs of individuals within the same cluster. This ensures that the resulting `self-similarity' metric $d_{ll}$ is properly scaled by the cluster’s internal dispersion. Consequently, $d_{ll}$ acts as a baseline of internal coherence against which cross-cluster distances are compared.
    
    As intended, $d_{lj} \leq d_{ll}$ for all $l, j \in \{1, \dots, J\}$. The parameter $\chi$ scales the rate of decay for cluster similarity. Larger values of $\chi$ facilitate global pooling across the cluster network, whereas as $\chi \to 0$, the model recovers local identification by restricting borrowing to identical units. We investigate sensitivity to $\chi$ and $\zeta$ (the relative weight of cluster vs. individual distances) in Section~\ref{SWBB.simulation}.
    
    We then specify a Dirichlet prior for $\omega_{j \mid \left( l \right)} = \mP[\bbc \in \Sc[j] \mid \bbv \in \Sv[l] ]$, $\bbomega_{\left( l \right)} \sim \mdir{\alpha_{l}^{\omega} \bbomega_{\left( l ; 0 \right)}}$ where we set $\bbomega_{\left( l ; 0 \right)} = \boldsymbol{\ndist}_{l} = \left( \ndist_{l1}, \ldots, \ndist_{lJ} \right)$ and $\alpha_{l}^{\omega} > 0$, which makes the expected value of $\omega_{j \mid \left( l \right)}$ proportional to the distance $\ndist_{lj}$. This is a conjugate model with posterior $\bbomega_{\left( l \right)} \mid \Sc, \Sv[l] \sim \mdir{\bbomega_{\left( l ; * \right)}}$, where $\bbomega_{\left( l ; * \right)} = \left( \alpha_{l}^{\bbomega} \ndist_{l1} + \delta_{l1}, \ldots, \alpha_{l}^{\bbomega} \ndist_{lJ} + \delta_{lJ} \right)$ and $\delta_{lj} = I \left( l = j \right)$:
    \begin{align*}
        \mP[\bbomega_{\left( l \right)} \mid \Sc, \Sv[l] ]
        & \propto 
        \left\{ \mP[\bbc \in \bbC_{l} \mid \bbv \in \Sv[l] ; \bbomega_{\left( l \right)} ] \right\} 
        \mP[\bbomega_{\left( l \right)} ; \alpha_{l}^{\omega}, \boldsymbol{\ndist}_{l} ] \\
        & \propto 
        \left\{ \omega_{l}^{\left( l \right)} \right\} 
        \left\{ \frac{\Gamma \left( \sum_{j=1}^{J} \alpha_{l}^{\omega} \ndist_{lj} \right)}{\prod_{j=1}^{J} \Gamma \left( \alpha_{l}^{\omega} \ndist_{lj} \right) } \prod_{j=1}^{J} \left( \omega_{j \mid \left( l \right)} \right)^{\alpha_{l}^{\omega} \ndist_{lj} - 1} \right\} \\
        & \propto 
        \left\{ \prod_{j=1}^{J} \left( \omega_{j \mid \left( l \right)} \right)^{\delta_{lj}} \right\} 
        \left\{ \prod_{j=1}^{J} \left( \omega_{j \mid \left( l \right)} \right)^{\alpha_{l}^{\omega} \ndist_{lj} - 1} \right\} \\
        & \propto 
        \left\{ \prod_{j=1}^{J} \left( \omega_{j \mid \left( l \right)} \right)^{\alpha_{l}^{\omega} \ndist_{lj} + \delta_{lj} - 1} \right\}.
    \end{align*}
    Finally, substituting the posterior parameters into \eqref{SWBB.swbb.eq.1}, the joint distribution is fully specified as:
    \begin{align*}
        \mP[\bbC = \bbc_{ji}, \bbV = \bbv_{l}]
        &= \mP[\bbV = \bbv_{l}]
        \mP[\bbC \in \Sc[j] \mid \bbV = \bbv_{l}]
        \mP[\bbC = \bbc_{ji} \mid \bbC \in \Sc[j]]
        \pi_{i \mid \left( j \right)} \omega_{j \mid \left( l \right)} \rho_{l}. 
    \end{align*}
    By construction, $\sum_{l=1}^{J} \sum_{j=1}^{J} \sum_{i=1}^{n_{j}} \mP[\bbC = \bbc_{ji}, \bbV = \bbv_{l} ] = 1$, as the weight vectors $\bbrho$, $\bbomega_{(l)}$, and $\bbpi_{(j)}$ each lie on their respective unit simplexes.

    Our approach assumes 
    \begin{align*}
        \mP[\bbc \in \Sc[j] \mid \bbv \in \Sv[l] ] =
        \begin{cases}
            \frac{\alpha_{l}^{\bbomega} \ndist_{lj}}{\alpha_{l}^{\bbomega} \ndist_{l \cdot} + 1} 
            \; & \text{if $j \ne l$} \\
            \frac{\alpha_{l}^{\bbomega} \ndist_{lj} + 1}{\alpha_{l}^{\bbomega} \ndist_{l \cdot} + 1} 
            \; & \text{if $j = l$} .
        \end{cases}
    \end{align*}
    Let $\eta_{lj}^{\bbomega} = \alpha_{l}^{\bbomega} \ndist_{lj}$, a quantity that can be interpreted as the addition of $\eta_{lj}^{\bbomega}$ `pseudo-subjects' from cluster $l$ to cluster $j$. This mechanism effectively facilitates information borrowing by incorporating more pseudo-subjects from clusters that are `closer' in the covariate space, as defined by $\ndist_{lj}$.This approach builds upon the HBB introduced by \citet{oganisian2022hierarchical}, where a similar weight is defined as $\eta_{lj}^{\nhbb} = \frac{n_{j}}{N} \alpha_{l}^{\nhbb} = \frac{n_{j}}{n_{l}} \tau^{\nhbb}$. However, a key limitation of the HBB is that its weights depend strictly on the relative sample sizes of the clusters, without accounting for their empirical similarity. 
    
    By decoupling the information-sharing weight from sample size and anchoring it to the distance metric $\ndist_{lj}$, the SWBB provides a more robust framework for causal identification in the presence of cluster-level heterogeneity.Similar to the HBB, we specify the concentration parameter as $\alpha_{l}^{\bbomega} = \frac{N}{n_{l}} \tau^{\bbomega}$, where $\tau^{\bbomega}$ represents the desired minimum sample size for the `closest' cluster (here, we set $\tau^{\bbomega} = 1$). Consequently, the weight $\eta_{lj}^{\bbomega} = \frac{N}{n_{l}} \ndist_{lj} \tau^{\bbomega}$ ensures that more pseudo-subjects are added as the sample size of cluster $l$ decreases, particularly from clusters $j$ that exhibit high similarity to cluster $l$. This allows for adaptive borrowing: if $\ndist_{lj} > \ndist_{lj'}$, cluster $l$ borrows more information from cluster $j$ than from cluster $j'$. Further details on the theoretical connection between SWBB and HBB are provided in Section~\ref{SWBB.supp.connection} of the Supplementary Materials.

    \input{figure/figureSWBB_borrowingcomparison}
    We provide a conceptual comparison of the information-sharing mechanisms across different Bayesian Bootstrap specifications in Figure~\ref{SWBB.borrowingcomparison}. In the standard BB, clusters are treated as independent units, with no information shared across cluster boundaries. The HBB \citep{oganisian2022hierarchical} extends this by allowing bi-directional borrowing, though the weights are determined strictly by the relative sample sizes of the clusters. In contrast, our proposed SWBB framework adaptively weights the borrowing strength based on both cluster size and the empirical similarity (covariate distance $d_{lj}$) between clusters. As illustrated by the varying edge thicknesses and styles in Figure~\ref{SWBB.borrowingcomparison}, SWBB decouples borrowing strength from mere sample size; it facilitates aggressive borrowing (solid, thick lines) from similar clusters while effectively discounting information (dashed, thin lines) from dissimilar ones, regardless of their size. For example, if the distance between cluster $J_2$ and $J_4$ is large, indicating they are dissimilar, the model will effectively discount their shared information—represented by dashed, thin lines—even if the clusters are large. Conversely, `closer' clusters like $J_2$ and $J_3$ will exhibit stronger borrowing, represented by solid, thick lines.

\section{Models for outcome and mediator} \label{SWBB.outcome}
    Suppose we observe $N$ independent subjects with data $\{ Y_{ji}, M_{j}, Z_{j}, \bbV_{j}, \bbC_{ji} \}_{i \in 1:n_{j}}^{j \in 1:J}$, where $n_{j}$ denotes the size of cluster $j$ and $N = \sum_{j} n_{j}$. While our framework is broadly compatible with various modeling strategies—including nonparametric approaches such as Bayesian Additive Regression Trees (BART) or Dirichlet Process (DP) mixture models—we utilize parametric regression models here to characterize the individual-level outcome and cluster-level mediator. This choice is motivated by two primary considerations. First, the use of parametric specifications allows us to isolate and evaluate the performance of the SWBB more clearly. Second, given the relatively small number of clusters ($J=24$) and the cluster-level nature of the mediator, parametric models provide the necessary stability for posterior estimation where data-intensive nonparametric methods might struggle. Across all analyses in this application, we evaluate and compare the posterior distributions generated by three Bayesian bootstrap variants for the confounder distribution: the standard BB, the HBB, and our proposed SWBB.

    The proposed framework is adaptable to diverse data types for both the outcome and mediator, including continuous, categorical, or ordinal variables, by selecting appropriate link functions $g_{Y}(\cdot)$ and $g_{M}(\cdot)$. The models are specified as:
    \begin{gather*}
        \mE[Y_{ji} \mid M_{j}, Z_{j}, \bbV_{j}, \bbC_{ji} ; \psi_{j}, \bbtheta^{y}] 
        = g_{Y}^{-1} \left( \mathbb{X}_{ji}^{y} \bbtheta^{y} + \psi_{j} \right) \\
        \mE[M_{j} \mid Z_{j} , \bbV_{j}, \overline{\bbC}_{j} ; \bbtheta^{m}]
        = g_{M}^{-1} \left( \mathbb{X}_{j}^{m} \bbtheta^{m} \right),
    \end{gather*}
    where the design matrices are defined as $\mathbb{X}_{ji}^{y} = (1, M_{j}, Z_{j}, \bbV_{j}, \bbC_{ji})$ and $\mathbb{X}_{j}^{m} = (1, Z_{j}, \bbV_{j}, \overline{\bbC}_{j})$. The vectors $\bbtheta^{y}$ and $\bbtheta^{m}$ represent the regression coefficients: $\bbtheta^{y} = (\theta_{0}^{y}, \theta_{m}^{y}, \theta_{z}^{y}, \bbtheta_{v}^{y}, \bbtheta_{c}^{y})^\top$ and $\bbtheta^{m} = (\theta_{0}^{m}, \theta_{z}^{m}, \bbtheta_{v}^{m}, \bbtheta_{\bar{c}}^{m})^\top$. For the outcome model, we assume a cluster-level random effect $\psi_{j} \sim \mathcal{N}(0, \sigma_{\psi}^2)$. 

\section{Posterior computation for causal mediation} \label{SWBB.posterior}
    We perform posterior inference by integrating the parametric outcome and mediator models over the non-parametric SWBB distribution of confounders. Specifically, at each MCMC iteration $d = 1, \dots, D$, we:

    \begin{enumerate}
        \item Sample Confounder Weights:
            \begin{enumerate}
                \item Draw cluster-level weights $\bbrho = (\rho_1, \dots, \rho_{J}) \sim \text{Dir}(\bbone_{J})$, defining the marginal distribution $\mP[\bbV = \bbv_l] = \rho_l$.
                
                \item Draw inter-cluster borrowing weights $\bbomega_{(l)} \sim \text{Dir}(\bbomega_{(l;*)})$, defining the conditional probability $\mP[\bbC \in \Sc[j] \mid \bbV = \bbv_l] = \omega_{j \mid l}$.
                
                \item Draw intra-cluster weights $\bbpi_{(j)} \sim \text{Dir}(\bbone_{n_{j}})$, defining the probability of selecting subject $i$ within cluster $j$: $\mP[\bbC = \bbc_{ji} \mid \bbC \in \Sc[j]] = \pi_{i \mid j}$.
                
            \end{enumerate}
        
        \item Sample Potential Mediator: Given posterior draws $(\bbtheta, \bbpsi)$ for the outcome and mediator models, we generate the potential mediator $m'$ from the predictive distribution $\mP[M (z') \mid Z_ = z', \overline{\bbC} = \overline{\bbc}, \bbV = \bbv_{l}; \bbtheta ]$.

        \item Compute Level-Specific Estimands: We obtain the mediation effects by marginalizing over the reconstructed joint PMF: $\mP[\bbC = \bbc_{ji}, \bbV = \bbv_l] = \rho_l \omega_{j \mid l} \pi_{i \mid j}$.
        \begin{itemize}
            \item \textit{Population-level mediation}; Marginalize over $\left\{ \bbC, \bbV \right\}$ pairs;
            \begin{align*}
                \mE[Y ( z, M (z') )] 
                & \approx 
                \sum_{l=1}^{J} \rho_{l} \sum_{j=1}^{J} \omega_{j \mid \left( l \right)} \sum_{i=1}^{n_{j}} \pi_{i \mid \left( j \right)} \mE[Y \mid M = m', Z = z, \bbC = \bbc_{ji}, \bbV = \bbv_{l}; \bbtheta^{y}, \psi^{y}]
            \end{align*}
            
            \item \textit{Cluster-level mediation}; Marginalize over $\left\{ \bbC \right\}$ for a given $\bbV$;
            \begin{align*}
                \mE[Y ( z, M (z') ) \mid \bbV = \bbv_{l}]
                & \approx 
                \sum_{j=1}^{J} \omega_{j \mid \left( l \right)} \sum_{i=1}^{n_{j}} \pi_{i \mid \left( j \right)} \mE[Y \mid M = m', Z = z, \bbC = \bbc_{ji}, \bbV = \bbv_{l}; \bbtheta^{y}, \psi^{y}]
            \end{align*}
            
            \item \textit{Individual-level mediation}; Marginalize over $\left\{ \bbV \right\}$ for a given $\bbC$;
            \begin{align*}
                \mE[Y ( z, M (z') ) \mid \bbC = \bbc_{ji}] 
                & \approx 
                \frac{1}{\sum_{t=1}^{J} \sum_{s=1}^{J} \sum_{r=1}^{n_{s}} \rho_{t} \omega_{s \mid \left( t \right)} \pi_{r \mid \left( s \right)}} \sum_{l=1}^{J} \rho_{l} \omega_{j \mid \left( l \right)} \pi_{i \mid \left( j \right)} \\
                & \hspace{10em} \mE[Y \mid M = m', Z = z, \bbC = \bbc_{ji}, \bbV = \bbv_{l}; \bbtheta^{y}, \psi^{y}]
            \end{align*}

        \end{itemize}
    \end{enumerate}
    Repeating this procedure across $D$ draws yields the posterior distribution for each estimand. In both our simulation studies and the empirical data analysis, we implemented this Gibbs sampler by retaining 2,000 iterations after discarding an initial 2,000 iterations as burn-in. This framework ensures that while individual-level confounders from all clusters contribute to the cluster-level estimate for $\bbv_{l}$, their influence is adaptively weighted by the similarity-based weights $\omega_{j \mid l}$. See Section~\ref{SWBB.supp.conditional} of the Supplementary Materials for the mediation effects conditional on a subset of individual- or cluster-level confounders.

\section{Simulation} \label{SWBB.simulation}

\subsection{Simulation scenarios} \label{SWBB.simulation.scenarios} 
    To evaluate the performance of the SWBB, we conducted a simulation study across 1,000 datasets per scenario, focusing on the marginal NIE, NDE, and ATE. The design, partially inspired by \citet{oganisian2021bayesian}, utilizes $J=24$ clusters with $n_{j}=100$ ($N=2,400$) and a cluster-level treatment $Z_{j}$ to mirror a cluster-randomized trial. This configuration specifically mirrors the structure of the BIGPIC trial detailed in Section~\ref{SWBB.BIGPIC}, which also consists of $J=24$ clusters. This alignment ensures that our simulation results provide relevant insights into the model's performance in a realistic, small-sample CRT setting.

    We compared the standard BB and HBB against the SWBB using three distance function weight configurations: (a) individual-level only, (b) cluster-level only, and (c) a combination of both. All models utilized the parametric structures from Section~\ref{SWBB.outcome}—a mixed-effects model for outcome $Y$ and a fixed-effects model for mediator $M$—under the Sequential Ignorability (SI) assumption.

    \input{table/tableSWBB_sim_scn}

    The simulation incorporates three cluster-level ($\bbV$) and six individual-level ($\bbC$) confounders. As detailed in Table~\ref{SWBB:tab:sim:scn}, we systematically vary the cross-level dependency across three primary scenarios:
    \begin{itemize}
        \item \textbf{Scenario 1 (Independence):} $\bbV$ and $\bbC$ are independent ($\bbV \indep \bbC$), representing a baseline hierarchical structure where confounders at different levels do not interact.
        \item \textbf{Scenario 2 (Weak Dependence):} $\bbV$ and $\bbC$ exhibit a weak dependence structure, introducing a moderate degree of cross-level confounding between the cluster and individual levels.
        \item \textbf{Scenario 3 (Strong Dependence):} $\bbV$ and $\bbC$ exhibit a strong dependence structure, creating a challenging environment that reflects severe multi-level confounding.
    \end{itemize}

    These scenarios reflect real-world complexities like multi-level confounding and finite cluster sizes. We report performance in terms of bias, root mean squared error (RMSE), and 95\% nominal coverage probability.

    These configurations reflect real-world complexities such as multi-level confounding and finite cluster sizes. For each scenario, we report performance based on bias, root mean squared error (RMSE), and 95\% nominal coverage probability.

\subsection{Results} \label{SWBB.simulation.results}
    \input{table/tableSWBB_sim_results}
    Table~\ref{SWBB:tab:sim:results} presents the performance of the BB, HBB, and SWBB estimators using the parametric mixed-effects and fixed-effects models described in Section~\ref{SWBB.outcome}. The results indicate that all causal estimates—NIE, NDE, and ATE—are approximately unbiased across all scenarios. Overall, the findings demonstrate that the proposed structural modeling approach provides robust point estimation regardless of the dependence structure between $\bbV$ and $\bbC$.

    We observe that the RMSEs of the SWBB decrease as the value of $\chi$ increases when $\bbV$ and $\bbC$ are independent (Scenario 1). Specifically, in Scenario 1, the RMSE for the NIE under the SWBB ($\chi = 1, \zeta = 0.5$) is approximately 5\% lower than that of the standard BB (0.81 vs. 0.85). The precision gains are similarly evident for the ATE; in Scenario 1, the RMSE of the SWBB ($\chi = 1, \zeta = 0.5$) is roughly 4.4\% lower than that of the BB (0.87 vs. 0.91) and 3.3\% lower than the HBB (0.87 vs. 0.90).

    In scenarios where $\bbV$ and $\bbC$ are dependent (Scenarios 2 and 3), the RMSEs for the NIE under the SWBB decrease as $\chi$ increases up to $\chi = 1$, but remain stable thereafter. For instance, in Scenario 3, the RMSE for the NIE decreases from 0.85 at $\chi = 0.01$ to 0.81 at $\chi = 1$, representing a 4.7\% improvement in precision. This suggests that while borrowing information across clusters significantly improves precision, the gains plateau once the optimal balance of cross-cluster information is reached, depending on the strength of the dependence between $\bbV$ and $\bbC$.

    A notable pattern across all scenarios is that the precision gains of the SWBB are prominent for the NIE and ATE, whereas the RMSE for the NDE remains relatively stable across all bootstrap specifications. This differential impact is structurally tied to the parametric formulation of the outcome model and the definition of the estimands. Under a generalized linear model with a random intercept, the direct effect acts largely as an intercept shift; consequently, marginalizing over the nonparametric confounder distribution $\mP[\bbC, \bbV]$ introduces minimal additional variance. However, if highly flexible models with complex exposure-confounder interactions (e.g., BART) were employed, this marginalization would carry more uncertainty, likely allowing the SWBB to improve NDE precision as well. Conversely, estimating the NIE requires predicting the potential mediator, which is highly sensitive to within-cluster data sparsity due to its reliance on aggregated individual confounders (e.g., $\overline{\bbC}_{j}$). By borrowing information across similar clusters, the SWBB stabilizes the conditional confounder distribution $\mP[\bbC \mid \bbV]$, directly yielding the substantial variance reduction observed in the NIE. Consequently, the efficiency gains in the ATE are almost entirely inherited from this stabilized NIE.

    Notably, the RMSEs under the SWBB are consistently lower than those under both the BB and HBB for $\chi \geq 1$, while maintaining a nominal coverage probability (CP) of approximately 0.95–0.97. This implies that the SWBB offers tighter uncertainty bounds and greater precision in posterior estimation without sacrificing inferential validity.

    As expected, the RMSEs of the standard BB are similar to those of the SWBB when $\chi$ is very small ($\chi = 0.01$). This aligns with the theoretical properties described in Section~\ref{SWBB.similarity}; as the decay rate $\chi \to 0$, cross-cluster borrowing is heavily penalized, and the SWBB naturally recovers the isolated, local identification of the standard BB. Furthermore, we note that changes in $\zeta$ do not have a significant impact, as it serves as a weight for the distance of cluster-level confounders $\bbV$ versus individual-level confounders $\bbC$. We generally recommend using $\zeta = 0.5$, though its selection may be application-specific. For the real data analysis, we proceed with $\chi = 1$ and $\zeta = 0.5$.

\section{BIGPIC study} \label{SWBB.BIGPIC}

\subsection{Study setting and analytic framework} \label{SWBB.BIGPIC.setting}
    We applied the proposed SWBB approach to data from the Bridging Income Generation with Group Integrated Care (BIGPIC) trial \citep{vedanthan2017bridging, vedanthan2021group}. The study was conducted in Western Kenya, where $J=24$ health facility sites (clusters) were randomly assigned to one of four treatment arms ($Z$): (1) usual care (UC, $Z=0$), (2) usual care plus microfinance (UC-MF, $Z=1$), (3) group medical visits (GMV, $Z=2$), and (4) group medical visits integrated into microfinance groups (GMV-MF, $Z=3$). The primary outcome ($Y$) is the one-year change in SBP.

    To satisfy the sequential ignorability assumptions required for causal identification, we controlled for a comprehensive set of baseline individual-level confounders ($\bbC$) and cluster-level factors ($\bbV$). The individual-level confounders included age ($C_{1}$), gender ($C_{2}$), livestock ownership ($C_{3}$), employment status ($C_{4}$), diabetes status ($C_{5}$), and cluster-centered baseline SBP ($C_{6}$). At the cluster level, we incorporated county ($V_{1}$), health facility type ($V_{2}$), and the baseline level of microfinance group penetration (GISE penetration) in the community ($V_{3}$). Furthermore, to account for potential cross-level confounding and contextual effects, we included the cluster-specific mean of baseline SBP ($V_{4}$) as well as the cluster-level means of all individual-level confounders ($\bar{C}_{1}$ through $\bar{C}_{6}$).
    
    To isolate the additive effects of the integrated care model, we focus on two distinct pairwise comparisons. First, we examine the additive effect of GMV given MF ($Z=1$ vs. $Z=3$) within the context of microfinance-based social networks (\textit{relation} = 1). Second, we evaluate the additive effect of MF given GMV ($Z=2$ vs. $Z=3$) within the group medical visit networks (\textit{relation} = 2). For each comparison, four sociometric network characteristics measured at 12 months—density, transitivity, cohesion, and average path length (APL)—are evaluated as potential mediators ($M$).
    
    While these network characteristics are theoretically interrelated, we evaluate each mediator in a single-mediator framework. Given the relatively small number of clusters ($J=24$), we do not have sufficient data to reliably support a multiple-mediator model, which would require significantly higher cluster-level degrees of freedom to achieve stable posterior estimation. Consequently, by analyzing each mediator separately, we maintain the robustness of the SWBB approach and ensure that the precision gains from our cross-cluster borrowing strategy are not offset by the over-parameterization of the mediator model.
    
\subsection{Results} \label{SWBB.BIGPIC.results}
    We applied the SWBB approach to the BIGPIC trial data to evaluate the mediating role of sociometric network characteristics on the change in SBP. As detailed in Section~\ref{SWBB.outcome}, our observed data model employs a parametric framework consisting of a mixed-effects model for the outcome ($Y$) to account for cluster-level correlation, alongside a fixed-effects model for the mediators ($M$).
    
\subsubsection{Comparison of MF given GMV-MF ($Z=1$ vs. $Z=3$)} \label{SWBB.BIGPIC.results.1vs3}
    Before conducting the causal mediation analysis, we summarized the primary outcome and network mediators for the analytic samples. For the comparison of UC-MF ($Z=0$ in analytic model) and GMV-MF ($Z=1$ in analytic model), the sample included 1,243 individuals. Both groups showed a mean reduction in SBP, with the integrated GMV-MF group exhibiting a slightly larger reduction (-16.4 mmHg, SD 22.8) compared to the UC-MF group (-15.7 mmHg, SD 20.9). Regarding network mediators, the UC-MF clusters exhibited higher mean density (0.0086 vs. 0.0053) and transitivity (0.123 vs. 0.045) compared to the integrated arm, while mean cohesion was 1.43 and 1.08, respectively. The mean APL was 0.738 for the UC-MF arm and 0.936 for the GMV-MF arm.
    
    \input{table/tableSWBB_BIGPIC_Z1vsZ3_m12}
    Table~\ref{SWBB:tab:Z1vsZ3_main} displays the estimates of NIE, NDE, and ATE for the comparison between usual care plus microfinance ($Z=1$) and integrated GMV-MF ($Z=3$). It should be noted that none of the estimated causal effects were statistically significant, as all 95\% credible intervals included zero across all methods and mediators. Nevertheless, the SWBB consistently produced the smallest Credible Interval Lengths (CIl) compared to the alternative bootstrap methods. On average, across all mediators in this comparison, the CIl of the SWBB for the ATE was approximately 21\% lower than that of the BB and 22\% lower than that of the HBB. For instance, in the Density mediator analysis, the CIl for the ATE under SWBB was 4.291, compared to 5.475 for BB and 5.509 for HBB. While the point estimates for the NIE often clustered around zero (e.g., SWBB Est: -0.179, SD: 0.928 for Density), the SWBB sharpened the posterior precision, providing tighter bounds and more stable evidence regarding the clinical impact of adding GMV to an existing microfinance framework.

    It was also of clinical interest to estimate causal effects conditional on a cluster-level covariate and individual-level covariate; computations to do this are provided in Section~\ref{SWBB.supp.conditional} of the Supplementary Materials.
    When conditioning on GISE penetration—comparing Light (Condition 1) to Heavy (Condition 2)—we observed that the CIl for these conditional effects were markedly larger than those of the population average. For the Density mediator, the ATE CIl was 6.100 for Light GISE and 6.335 for Heavy GISE, compared to 4.291 for the population average. This increase in interval length is expected, as the conditional estimates are derived from a subset of the total clusters, reducing the effective sample size and increasing the posterior uncertainty. Despite this, the SWBB maintained greater stability by borrowing strength from similar clusters via the similarity-weighted framework.
    Conditioning on diabetes status—No (Condition 3) vs. Yes (Condition 4)—revealed a similar trend. For all mediators, the CIl for the diabetic subgroup (Yes) was substantially wider than for the population average or the non-diabetic subgroup. For example, in the Density analysis, the SWBB produced an ATE CIl of 9.838 for the diabetic subgroup, nearly double the population average CIl. This reflects the limited degrees of freedom available in smaller individual-level strata, yet the SWBB approach ensured the precision gains from the similarity-weighted borrowing were not entirely lost to over-parameterization.
    
\subsubsection{Comparison of GMV given GMV-MF ($Z=2$ vs. $Z=3$)} \label{SWBB.BIGPIC.results.2vs3}
    For the comparison between GMV ($Z=0$ in analytic model) and GMV-MF ($Z=1$ in analytic model), the sample consisted of 1,304 individuals. The mean reduction in SBP was -15.0 mmHg (SD 21.2) for the GMV group and -16.4 mmHg (SD 22.8) for the GMV-MF group. Mediator profiles were more comparable in this subset: mean density was 0.0092 for GMV and 0.0055 for GMV-MF; transitivity was 0.047 and 0.031, respectively; mean cohesion was 1.12 for GMV and 1.08 for GMV-MF; and mean APL was 0.900 for GMV and 0.938 for GMV-MF.
    
    \input{table/tableSWBB_BIGPIC_Z2vsZ3_m12}
    Table~\ref{SWBB:tab:Z2vsZ3_main} presents the additive effect of microfinance within group medical visit networks ($Z=2$ vs. $Z=3$). As with the previous comparison, none of the estimated causal effects were statistically significant, with all 95\% credible intervals including zero across all methods and mediators. Despite this, the primary advantage of the SWBB is its consistent out-performance of both the standard BB and the HBB in terms of inferential precision, specifically regarding the width of the Credible Intervals (CIl). For the Average Total Effect (ATE), the SWBB achieved a CIl that was approximately 22\% narrower than the standard BB (e.g., 4.680 vs. 6.092 for APL) and 18\% narrower than the HBB (4.680 vs. 5.718 for APL). In the Transitivity mediator analysis, the SWBB produced an ATE CIl of 4.653, a substantial reduction in uncertainty compared to the BB (5.967) and HBB (5.634). Across all four mediators—Density, Transitivity, Cohesion, and APL—the SWBB consistently yielded the most concentrated posterior distributions for the Natural Indirect Effect (NIE) and Natural Direct Effect (NDE). These results demonstrate that the SWBB's cross-cluster borrowing strategy effectively minimizes the over-dispersion typically encountered in hierarchical data models.
    
    The efficiency gains of the SWBB remained evident in the conditional analyses. For the Density mediator, the CIl for the ATE under Light GISE (Condition 1) was 5.106, significantly more precise than the population-average BB (6.023) or HBB (5.644) estimates. Even under the more heterogeneous Heavy GISE (Condition 2), the SWBB maintained a CIl of 7.401, which remains competitive despite the increased variance associated with community-level penetration subgroups.
    The SWBB demonstrated its robustness when applied to individual-level subgroups, which often suffer from data sparsity in clustered designs. For individuals without diabetes (Condition 3), the SWBB achieved highly precise ATE CIl values (e.g., 5.575 for Density and 5.521 for APL). For the higher-risk diabetes subgroup (Condition 4), where point estimates showed larger effects, the SWBB produced a CIl of 9.847 for Density and 9.868 for APL. While individual-level conditioning naturally leads to wider intervals than population-level averages, the SWBB remains the most efficient framework for these specific profiles.
    
\section{Discussion} \label{SWBB.discussion}
    We propose a semiparametric Bayesian inference method for causal mediation analysis in CRTs. The proposed method incorporates a `distance' metric between clusters, enabling borrowing more information from `closer' clusters. The advantages of this approach include its flexibility in utilizing the BB without requiring specific parametric assumptions for confounder distributions. Furthermore, the framework's full conjugacy facilitates the integration of diverse outcome modeling strategies, ensuring compatibility across various specifications for both the outcome and mediator models. While parametric regression models are employed for the outcome and mediator to ensure stability in settings with a limited number of clusters, the SWBB itself functions as a flexible, non-parametric framework for modeling the joint distribution of confounders.
    
    In the BIGPIC study, the analysis did not find evidence for significant population-level natural indirect, natural direct, or average treatment effects across the network mediators. However, the application of the SWBB demonstrated that the approach can accurately estimate natural effects even with a small number of clusters. A key finding was that the SWBB is primarily focused on reducing variance rather than bias, consistently producing smaller Credible Interval Lengths (CIl) than both the standard BB and HBB. Across all mediators, the CIl of the SWBB was lower than that of the BB and HBB, reflecting significant precision gains from its adaptive information-borrowing mechanism. Furthermore, while CIl for conditional effects were expectedly larger due to reduced sub-sample sizes, the SWBB sharpened posterior precision and successfully avoided the over-dispersion often observed in the standard BB when applied to hierarchical data.

    Beyond the current parametric specifications for the outcome and mediator, incorporating robust nonparametric models like BART or transitioning toward a fully Bayesian nonparametric framework--utilizing Dirichlet process mixture models (and their variations) for a CRT (e.g., \citet{ohnishi2025bayesian})--would provide further robustness to model misspecification. However, a small number of clusters may present challenges when applying nonparametric approaches to the mediator model, as these methods typically require more data to achieve stable estimation at the cluster level. Despite this limitation, such an approach would circumvent the constraint of defining causal effects based solely on the observed values of the covariates, thereby facilitating more generalized inference across the target population.
    
    Future research will investigate the integration of `clever covariates' into the proposed model as discussed in \citet{linero2022mediation}. It is also important to develop sensitivity analyses for potential violations of the sequential ignorability assumptions. Additionally, expanding the causal mediation model from a single mediator to accommodate multiple or time-varying mediators holds promise for advancing the field. One potential avenue for addressing identification is to investigate weaker assumptions, such as the mixture of mediator induction equivalence assumptions proposed in prior work \citep{daniels2012bayesian, kim2017framework}. Furthermore, extending the SWBB to smoothed versions of the BB could be advantageous. This extension would involve using a point-mass empirical distribution with a parametric kernel to induce smoothness \citep{efron1983leisurely, silverman1987bootstrap, oganisian2022hierarchical}.

\section*{Acknowledgments}
    This work was supported by the following NIH R01 grants: CA183854, HL158963, and HL166324.

\section*{Supplementary Materials}
    The source code is available at \url{https://github.com/WoojungBae/SWBB}.

\section*{Data Availability}
    The data that support the findings in this paper will be shared on reasonable request to the corresponding author.
    
    {
    \setlength{\bibsep}{0pt}
    \bibliographystyle{agu}
    \bibliography{fileReference}      
    }
    
\clearpage
\renewcommand*{\thetable}{\Alph{table}}
\renewcommand*{\thefigure}{\Alph{figure}}
\renewcommand*{\thesection}{\Alph{section}}
\setcounter{table}{0}
\setcounter{figure}{0}
\setcounter{section}{0}

\section{Identification of causal mediation effects} \label{SWBB.supp.identification}
    The structural assumptions introduced in Section~\ref{SWBB.causal.identifying} of the main text are used to nonparametrically identify the nested potential outcome $\mE[Y ( z, M (z') ) \mid \bbC = \bbc, \bbV = \bbv ]$ from the observed data distribution. Before detailing the sequential derivation, we note that the no-interference component of the Cluster-level SUTVA \eqref{SWBB.causal.eq.1} is fundamentally invoked to define this localized potential outcome. Without \eqref{SWBB.causal.eq.1}, an individual's potential outcome would depend on the full vectors of assignments and mediators across all clusters ($\mathbf{Z}$ and $\mathbf{M}$). By assuming no cross-cluster interference, we mathematically reduce $Y_{ji}(\mathbf{Z}, \mathbf{M})$ to $Y (z, m)$, permitting the simplified scalar notation used in the expectation below. Applying the mediation formula \citep{imai2010identification}, the derivation proceeds as follows:
    \begin{align*}
        & \mE[ Y ( z, M (z') ) \mid \bbC = \bbc, \bbV = \bbv] \\
        & = \int \mE[ Y (z, m') \mid M (z') = m', Z = z', \bbC = \bbc, \bbV = \bbv] \ndF_{M (z') \mid Z = z', \bbC = \bbc, \bbV = \bbv} \left( m' \right) \\
        & = \int \mE[ Y (z, m') \mid Z = z', \bbC = \bbc, \bbV = \bbv] \ndF_{M (z') \mid Z = z', \bbC = \bbc, \bbV = \bbv} \left( m' \right) && \mbox{by \eqref{SWBB.causal.eq.4}} \\
        & = \int \mE[ Y (z, m') \mid Z = z, \bbC = \bbc, \bbV = \bbv] \ndF_{M (z') \mid Z = z', \bbC = \bbc, \bbV = \bbv} \left( m' \right) && \mbox{by \eqref{SWBB.causal.eq.3}} \\
        & = \int \mE[ Y (z, m') \mid M (z) = m', Z = z, \bbC = \bbc, \bbV = \bbv] \ndF_{M (z') \mid Z = z', \bbC = \bbc, \bbV = \bbv} \left( m' \right) && \mbox{by \eqref{SWBB.causal.eq.4}} \\
        & = \int \mE[ Y \mid M = m', Z = z, \bbC = \bbc, \bbV = \bbv] \ndF_{M \mid Z = z', \bbC = \bbc, \bbV = \bbv} \left( m' \right), && \mbox{by \eqref{SWBB.causal.eq.2}}
    \end{align*}
    where each step in the sequential derivation is directly justified by our core causal assumptions. First, we drop the conditioning on the potential mediator $M (z')$ using the mediator ignorability assumption \eqref{SWBB.causal.eq.4}, since $Y (z, m') \indep M (z') \mid Z=z'$. Next, we apply the treatment ignorability assumption \eqref{SWBB.causal.eq.3} to exchange the conditioning on the treatment assignment from $Z = z'$ to $Z = z$. We then apply mediator ignorability \eqref{SWBB.causal.eq.4} again to safely condition on $M (z) = m'$. Finally, the last equality invokes the consistency rule of the Cluster-level SUTVA \eqref{SWBB.causal.eq.2}; because we condition on $Z = z$ and $M (z) = m'$, the potential outcome $Y (z, m')$ corresponds precisely to the observed outcome $Y$, and similarly, the potential mediator $M (z')$ within the integration measure maps directly to the observed mediator $M$ since we are conditioning on $Z = z'$.

    Having established this nonparametric identification entirely in terms of observed variables, we now adapt it to accommodate the specific modeling framework introduced in Section~\ref{SWBB.outcome}. Because our generalized linear model for the outcome incorporates a cluster-level random intercept $\bbpsi$ to account for within-cluster correlation, the expectation must be further marginalized over the distribution of this random effect. Thus, the final identification formula expands to:
    \begin{align*}
        & \mE[ Y ( z, M (z') ) \mid \bbC = \bbc, \bbV = \bbv] \\ 
        & = \int \mE[ Y \mid M = m', Z = z, \bbC = \bbc, \bbV = \bbv] \ndF_{M \mid Z = z', \bbC = \bbc, \bbV = \bbv} \left( m' \right) \\
        & = \iint \mE[ Y \mid M = m', Z = z, \bbC = \bbc, \bbV = \bbv ; \bbpsi] \ndF_{M \mid Z = z', \bbC = \bbc, \bbV = \bbv } \left( m' \right) \ndF_{\bbpsi} \left( \bbpsi \right).
    \end{align*}

\section{Conditional mediation effects} \label{SWBB.supp.conditional}
    $\bullet$ Let $\Sv[l][q] = \left\{ t: \bbV_{t}^{q} = \bbv_{l}^{q}, t = 1, \ldots, J \right\}$. Then.
    \begin{align*}
        \mP[ \bbV_{l}^{q} ] 
        = 
        \sum_{t \in \Sv[l][q]} \sum_{s=1}^{J} \sum_{r=1}^{n_{s}} \mP[ \bbC_{sr}, \bbV_{t}^{-q}, \bbV_{l}^{q} ] 
        = 
        \sum_{t \in \Sv[l][q]} \sum_{s=1}^{J} \sum_{r=1}^{n_{s}} \pi_{r \mid \left( s \right)} \omega_{s \mid \left( t \right)} \rho_{t} 
        = 
        \sum_{t \in \Sv[l][q]} \sum_{j=1}^{J} \omega_{j \mid \left( t \right)} \rho_{t} 
        = 
        \sum_{t \in \Sv[l][q]} \rho_{t},
    \end{align*}
    and
    \begin{align*}
        \mP[ \bbC_{ji}, \bbV_{t}^{-q} \mid \bbV_{l}^{q} ]
        = \frac{\mP[ \bbC_{ji}, \bbV_{t}^{-q}, \bbV_{l}^{q} ]}{\mP[ \bbV_{l}^{q} ]} 
        = \frac{\mP[ \bbC_{ji}, \bbV_{t} ]}{\mP[ \bbV_{l}^{q} ]} 
        = \frac{\pi_{i \mid \left( j \right)} \omega_{j \mid \left( t \right)} \rho_{t}}{\sum_{t \in \Sv[l][q]} \rho_{t}}
        \overset{\text{set}}{=} \lambda_{tji}^{v_{l}^{q}}
    \end{align*}
    where $j = 1, \ldots, J$, $i = 1, \ldots, n_{j}$ and $t \in \Sv[l][q]$. Note that
    \begin{align*}
        \sum_{t \in \Sv[l][q]} \sum_{j=1}^{J} \sum_{i=1}^{n_{j}} \mP[ \bbC_{ji}, \bbV_{t}^{-q} \mid \bbV_{l}^{q} ]
        = \sum_{t \in \Sv[l][q]} \sum_{j=1}^{J} \sum_{i=1}^{n_{j}} \lambda_{tji}^{v_{l}^{q}}
        = \sum_{t \in \Sv[l][q]} \sum_{j=1}^{J} \sum_{i=1}^{n_{j}} \frac{\pi_{i \mid \left( j \right)} \omega_{j \mid \left( t \right)} \rho_{t}}{\sum_{u \in \Sv[l][q]} \rho_{u}}
        = \frac{\sum_{t \in \Sv[l][q]} \rho_{t}}{\sum_{u \in \Sv[l][q]} \rho_{u}}
        = 1.
    \end{align*}
    $\Rightarrow$ \textit{Cluster-level mediation}; Marginalize over $\left\{ \bbC, \bbV^{-q} \right\}$ for a given $\bbV^{q}$
    \begin{align*}
        & \mE[Y ( z, M (z') ) \mid \bbV^{q} = \bbv_{l}^{q}] \\
        & = 
        \iiint \mE[ Y \mid M = m', Z = z, \bbC = \bbc , \bbV = \bbv ; \bbtheta^{y}, \psi^{y}] \\
        & \hspace{10em} \ndF_{M \mid Z = z', \bbC = \bbc, \bbV = \bbv ; \bbtheta^{m}, \psi^{m}} \left( m' \right) \ndF_{ \bbpsi } \left( \bbpsi \right) \ndF_{\bbC, \bbV^{-q} \mid \bbV^{q} = \bbv^{q}} \left( \bbc, \bbv^{-q} \right) \\
        & = 
        \sum_{t \in \Sv[l][q]} \sum_{j=1}^{J} \sum_{i=1}^{n_{j}} \lambda_{j,i,t}^{v_{l}^{q}} \mE[Y \mid M = m', Z = z, \bbC = \bbc_{ji}, \bbV^{-q} = \bbv_{t}^{-q}, \bbV^{q} = \bbv_{l}^{q}; \bbtheta^{y}, \psi^{y}]
    \end{align*}

    $\bullet$ Let $\Sc[ji][p] = \left\{ \left( s, r \right): \bbC_{sr}^{p} = \bbc_{ji}^{p}, s = 1, \ldots, J, r = 1, \ldots, n_{s} \right\}$. Then,
    \begin{align*}
        \mP[ \bbC_{ji}^{p} ]
        = 
        \sum_{l=1}^{J} \mathop{\sum\sum}_{\left( s, r \right) \in \Sc[ji][p]} \mP[ \bbC_{sr}^{-p}, \bbC_{ji}^{p}, \bbV_{l} ] 
        = 
        \sum_{l=1}^{J} \mathop{\sum\sum}_{\left( s, r \right) \in \Sc[ji][p]} \pi_{r \mid \left( s \right)} \omega_{s \mid \left( l \right)} \rho_{l}
    \end{align*}
    and
    \begin{align*}
        \mP[ \bbC_{sr}^{-p}, \bbV_{l} \mid \bbC_{ji}^{p} ]
        = \frac{\mP[ \bbC_{sr}^{-p}, \bbC_{ji}^{p}, \bbV_{l} ]}{\mP[ \bbC_{ji}^{p} ]} 
        = \frac{\mP[ \bbC_{sr}, \bbV_{l} ]}{\mP[ \bbC_{ji}^{p} ]} 
        = \frac{\pi_{r \mid \left( s \right)} \omega_{s \mid \left( l \right)} \rho_{l}}{\sum_{t=1}^{J} \mathop{\sum\sum}_{\left( u, h \right) \in \Sc[ji][p]} \pi_{h \mid \left( u \right)} \omega_{u \mid \left( t \right)} \rho_{t}}
        \overset{\text{set}}{=} \lambda_{lsr}^{\bbc_{ji}^{p}}
    \end{align*}
    where for $\left( s, r \right) \in \Sc[ji][p]$ and $l = 1, \ldots, J$. Note that
    \begin{align*}
        \sum_{l=1}^{J} \mathop{\sum\sum}_{\left( s, r \right) \in \Sc[ji][p]} \mP[ \bbC_{sr}^{-p}, \bbV_{l} \mid \bbC_{ji}^{p} ]
        = \sum_{l=1}^{J} \mathop{\sum\sum}_{\left( s, r \right) \in \Sc[ji][p]} \lambda_{lsr}^{\bbc_{ji}^{p}}
        = \sum_{l=1}^{J} \mathop{\sum\sum}_{\left( s, r \right) \in \Sc[ji][p]} \frac{\pi_{r \mid \left( s \right)} \omega_{s \mid \left( l \right)} \rho_{l}}{\sum_{t=1}^{J} \mathop{\sum\sum}_{\left( u, h \right) \in \Sc[ji][p]} \pi_{h \mid \left( u \right)} \omega_{u \mid \left( t \right)} \rho_{t}}
        = 1.
    \end{align*}
    $\Rightarrow$ \textit{Individual-level mediation}; Marginalize over $\left\{ \bbC^{-p}, \bbV \right\}$ for a given $\bbC^{p}$ 
    \begin{align*}
        & \mE[Y ( z, M (z') ) \mid \bbC^{p} = \bbc_{ji}^{p}] \\
        & = 
        \iiint \mE[ Y \mid M = m', Z = z, \bbC = \bbc , \bbV = \bbv_{j} ; \bbtheta^{y}, \psi^{y}]  \\
        & \hspace{10em} \ndF_{M \mid Z = z', \bbC = \bbc, \bbV = \bbv ; \bbtheta^{m}, \psi^{m}} \left( m' \right) \ndF_{ \bbpsi } \left( \bbpsi \right) \ndF_{\bbC^{-p} = \bbc^{-p}, \bbV \mid \bbC^{p} = \bbc^{p}} \left( \bbc^{-p}, \bbv \right) \\
        & = 
        \sum_{l=1}^{J} \mathop{\sum\sum}_{\left( s, r \right) \in \Sc[ji][p]} \lambda_{s,r,l}^{\bbc_{ji}^{p}} \mE[Y \mid M = m', Z = z, \bbC^{-p} = \bbc_{sr}^{-p}, \bbC^{p} = \bbc_{ji}^{p}, \bbV = \bbv_{l}; \bbtheta^{y}, \psi^{y}].
    \end{align*}

\section{Connection to HBB} \label{SWBB.supp.connection}
    In this section, we provide a rigorous comparison between the SWBB and the HBB proposed by \citet{oganisian2022hierarchical}. We demonstrate that the HBB represents a specific case where information borrowing is governed exclusively by relative sample sizes, whereas the SWBB generalizes this mechanism by incorporating empirical covariate similarity.
    
    The HBB assumes that the conditional distribution of individual-level confounders follows:
    \begin{align*}
        \mP[\bbc \in \Sc[ji] \mid \bbv \in \Sv[l] ] =
        \begin{cases}
            \frac{\frac{1}{N} \alpha_{l}^{\nhbb}}{\alpha_{l}^{\nhbb} + n_{l}} \; & \text{if $j \ne l$} \\
            \frac{\frac{1}{N} \alpha_{l}^{\nhbb} + 1}{\alpha_{l}^{\nhbb} + n_{l}} \; & \text{if $j = l$} 
        \end{cases}
        \Rightarrow
        \mP[\bbc \in \Sc[j] \mid \bbv \in \Sv[l] ] =
        \begin{cases}
            \frac{\frac{n_{j}}{N} \alpha_{l}^{\nhbb}}{\alpha_{l}^{\nhbb} + n_{l}} \; & \text{if $j \ne l$} \\
            \frac{\frac{n_{l}}{N} \alpha_{l}^{\nhbb} + n_{l}}{\alpha_{l}^{\nhbb} + n_{l}} \; & \text{if $j = l$}.
        \end{cases}
    \end{align*}

    We can interpret $\eta_{lj}^{\nhbb} = \frac{n_{j}}{N} \alpha_{l}^{\nhbb} > 0$ as adding an additional $\eta_{lj}^{\nhbb}$ `pseudo-subjects' from the cluster $l$ to the cluster $j$. Higher $\alpha_{l}^{\nhbb}$ places more weight on the `pseudo-subjects' - who may have values unseen in cluster $\bbv \in \Sv[l]$ (\textit{i.e.}, more shrinkage towards the marginal). \cite{oganisian2022hierarchical} set $\alpha_{l}^{\nhbb} = \frac{N}{n_{l}} \tau^{\nhbb}$ where $\tau^{\nhbb} > 0$ is user-specified and can be interpreted as the minimum desired sample size in each cluster. For their simulations and applications, a value of $\tau^{\nhbb} = 100$ is used. Then, $\eta_{lj}^{\nhbb} = \frac{n_{j}}{N} \alpha_{l}^{\nhbb} = \frac{n_{j}}{N} \frac{N}{n_{l}} \tau^{\nhbb} = \frac{n_{j}}{n_{l}} \tau^{\nhbb}$, which can be interpreted as the greater the ratio of the sample size between cluster $l$ and cluster $j$, the more `pseudo-subjects' are included. This solely relies on the sample size of each cluster, but does not reflect the similarity between the two clusters in terms of the individual covariates in each cluster. This serves as motivation for our approach.
    
    In contrast, SWBB framework explicitly assumes:
    \begin{align*}
        \mP[\bbc \in \Sc[j] \mid \bbv \in \Sv[l] ] =
        \begin{cases}
            \frac{\alpha_{l}^{\bbomega} \ndist_{lj}}{\alpha_{l}^{\bbomega} \ndist_{l \cdot} + 1} 
            \; & \text{if $j \ne l$} \\
            \frac{\alpha_{l}^{\bbomega} \ndist_{lj} + 1}{\alpha_{l}^{\bbomega} \ndist_{l \cdot} + 1} 
            \; & \text{if $j = l$} .
        \end{cases}
    \end{align*}
    Here, $\eta_{lj}^{\bbomega} = \alpha_{l}^{\bbomega} \ndist_{lj}$ incorporates pseudo-subjects based on the distance $\ndist_{lj}$, ensuring that `closer' clusters contribute more heavily to the estimation. These pseudo-subjects can take any observed value from the marginal distribution, effectively borrowing information across the covariate space. Following a logic similar to the HBB, we set $\alpha_{l}^{\bbomega} = \frac{N}{n_{l}} \tau^{\bbomega}$, where $\tau^{\bbomega}$ is the desired minimum sample size for the `closest' cluster (we use $\tau^{\bbomega} = 1$). Consequently, $\eta_{lj}^{\bbomega} = \frac{N}{n_{l}} \ndist_{lj} \tau^{\bbomega}$ implies that smaller clusters receive more pseudo-subjects. Furthermore, for a fixed cluster $l$, if $\ndist_{lj} < \ndist_{lj'}$ (indicating cluster $j$ is more similar to $l$ than $j'$), the SWBB borrows more information from cluster $j$. Finally, it should be noted that both $\tau^{\nhbb}$ and $\tau^{\bbomega}$ are not identifiable and remain user-specified.

\end{document}

%% file: figure/figureSWBB_DAG.tex
\begin{figure}[htbp] 
    \centering
    \begin{tikzpicture}
        [
            rect/.style={minimum size=15pt,rectangle,draw},
            circ/.style={minimum size=15pt,circle,draw},
            arro/.style={->,>=stealth',semithick},
            dottarro/.style={->,>=stealth',densely dotted},
            biarro/.style={<->,>=stealth',densely dotted},
        ]
        
        \node (V) [circ] at (-2, -2) {$V_{j}$};
        \node (C) [circ] at (2, -2) {$\pmb{C}_{ji}$};
        
        \node (Z) [circ] at (-6, 0) {$Z_{j}$};   
        \node (M) [circ] at (0, 2) {$M_{j}$};    
        \node (Y) [circ] at (6, 0) {$Y_{ji}$};   
        
        \draw[arro] (Z) -- (Y) node[pos=0.4, above] {(a)};
        \draw[arro] (Z) -- (M) node[midway, above left] {(b)};
        \draw[arro] (M) -- (Y) node[midway, above right] {(c)};
        \draw[biarro] (V) -- (C); 
        
        \node [draw, fit=(V) (C), inner sep=4pt] (VC-box) {};
        
        \begin{scope}[dottarro]
            \draw (VC-box.north) -- (M.south);
            \draw (VC-box.east) -- (Y.south west);
        \end{scope}
        
    \end{tikzpicture}
    \caption{\label{SWBB.causalstructure} Causal structure of the CRT showing the cluster-level mediator ($M_{j}$), individual-level outcome ($Y_{ji}$), and the correlation between cluster ($V_{j}$) and individual ($\pmb{C}_{ji}$) confounders. The solid arrows represent the causal effects of interest: (a) the natural direct effect of treatment $Z_{j}$ on outcome $Y_{ji}$; (b) the effect of treatment $Z_{j}$ on the mediator $M_{j}$; and (c) the effect of the mediator $M_{j}$ on outcome $Y_{ji}$. Dotted lines indicate confounding pathways.}
\end{figure}
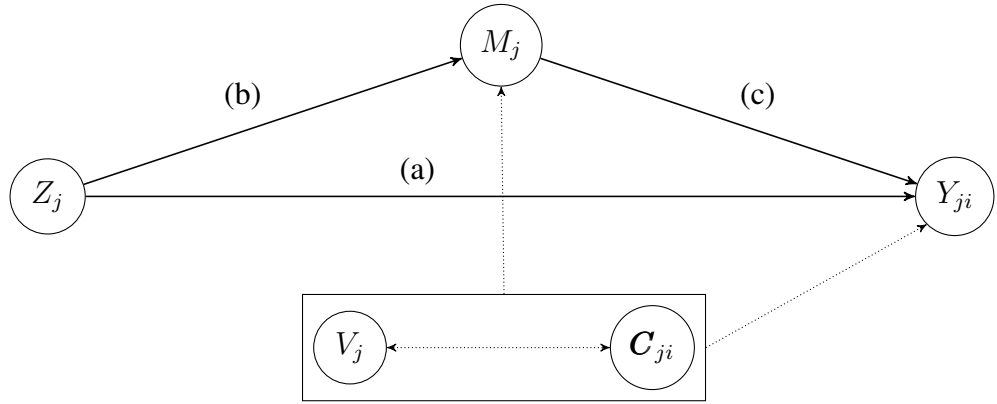

%% file: figure/figureSWBB_borrowingcomparison.tex
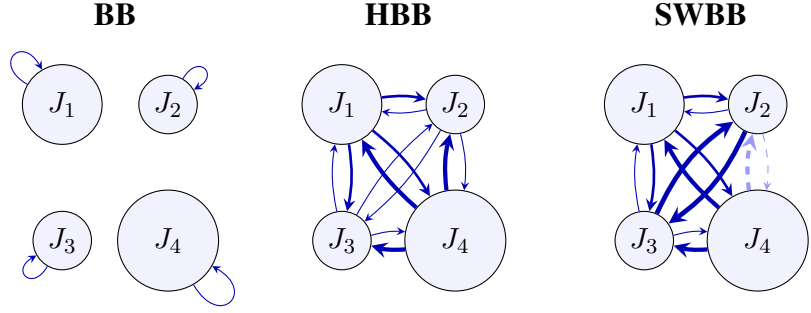
\begin{figure}[tbp]
    \centering
    \begin{tikzpicture}[
        base_circ/.style={circle, draw, fill=blue!5, inner sep=2pt, font=\small},
        size_j4/.style={base_circ, minimum size=38pt}, 
        size_j1/.style={base_circ, minimum size=30pt}, 
        size_j23/.style={base_circ, minimum size=22pt}, 
        line_j4/.style={line width=1.6pt},
        line_j1/.style={line width=1.0pt},
        line_j23/.style={line width=0.4pt},
        arro/.style={->, >=stealth, blue!70!black},
        dash_arro/.style={->, >=stealth, dashed, blue!40}
    ]

    \node (BB_label) at (-1.5, 1.2) {\textbf{BB}};
    \node (BB_j1) [size_j1] at (-2.2, 0) {$J_1$};
    \node (BB_j2) [size_j23] at (-0.8, 0) {$J_2$};
    \node (BB_j3) [size_j23] at (-2.2, -1.8) {$J_3$};
    \node (BB_j4) [size_j4] at (-0.8, -1.8) {$J_4$};
    
    \draw[arro, line width=0.4pt] (BB_j1) to [out=150,in=120,looseness=5] (BB_j1);
    \draw[arro, line width=0.4pt] (BB_j2) to [out=60,in=30,looseness=5] (BB_j2);
    \draw[arro, line width=0.4pt] (BB_j3) to [out=240,in=210,looseness=5] (BB_j3);
    \draw[arro, line width=0.4pt] (BB_j4) to [out=300,in=330,looseness=5] (BB_j4);

    \node (HBB_label) at (2.25, 1.2) {\textbf{HBB}};
    \node (HBB_j1) [size_j1] at (1.5, 0) {$J_1$};
    \node (HBB_j2) [size_j23] at (3, 0) {$J_2$};
    \node (HBB_j3) [size_j23] at (1.5, -1.8) {$J_3$};
    \node (HBB_j4) [size_j4] at (3, -1.8) {$J_4$};
    
    \draw[arro, line_j1] (HBB_j1) edge[bend left=10] (HBB_j4);
    \draw[arro, line_j4] (HBB_j4) edge[bend left=10] (HBB_j1);
    \draw[arro, line_j23] (HBB_j2) edge[bend left=10] (HBB_j3);
    \draw[arro, line_j23] (HBB_j3) edge[bend left=10] (HBB_j2);
    \draw[arro, line_j1] (HBB_j1) edge[bend left=10] (HBB_j2);
    \draw[arro, line_j23] (HBB_j2) edge[bend left=10] (HBB_j1);
    \draw[arro, line_j23] (HBB_j3) edge[bend left=10] (HBB_j4);
    \draw[arro, line_j4] (HBB_j4) edge[bend left=10] (HBB_j3);
    \draw[arro, line_j1] (HBB_j1) edge[bend left=10] (HBB_j3);
    \draw[arro, line_j23] (HBB_j3) edge[bend left=10] (HBB_j1);
    \draw[arro, line_j23] (HBB_j2) edge[bend left=10] (HBB_j4);
    \draw[arro, line_j4] (HBB_j4) edge[bend left=10] (HBB_j2);

    \node (SWBB_label) at (6.25, 1.2) {\textbf{SWBB}};
    \node (SWBB_j1) [size_j1] at (5.5, 0) {$J_1$};
    \node (SWBB_j2) [size_j23] at (7, 0) {$J_2$};
    \node (SWBB_j3) [size_j23] at (5.5, -1.8) {$J_3$};
    \node (SWBB_j4) [size_j4] at (7, -1.8) {$J_4$};

    \draw[arro, line width=1.8pt] (SWBB_j2) edge[bend left=15] (SWBB_j3);
    \draw[arro, line width=1.8pt] (SWBB_j3) edge[bend left=15] (SWBB_j2);

    \draw[dash_arro, line_j23] (SWBB_j2) edge[bend left=10] (SWBB_j4);
    \draw[dash_arro, line_j4] (SWBB_j4) edge[bend left=10] (SWBB_j2);

    \draw[arro, line_j1] (SWBB_j1) edge[bend left=10] (SWBB_j4);
    \draw[arro, line_j4] (SWBB_j4) edge[bend left=10] (SWBB_j1);
    \draw[arro, line_j1] (SWBB_j1) edge[bend left=10] (SWBB_j2);
    \draw[arro, line_j23] (SWBB_j2) edge[bend left=10] (SWBB_j1);
    \draw[arro, line_j1] (SWBB_j1) edge[bend left=10] (SWBB_j3);
    \draw[arro, line_j23] (SWBB_j3) edge[bend left=10] (SWBB_j1);
    \draw[arro, line_j4] (SWBB_j4) edge[bend left=10] (SWBB_j3);
    \draw[arro, line_j23] (SWBB_j3) edge[bend left=10] (SWBB_j4);

    \end{tikzpicture}
    \caption{\label{SWBB.borrowingcomparison} Conceptual comparison of information sharing across Bayesian Bootstrap specifications. Node sizes reflect relative subject counts ($n_j$) where $J_4 > J_1 > $ $J_3 = J_2$. In BB, clusters are independent. In HBB, all clusters share information bi-directionally with weights determined strictly by sample size. In SWBB, all clusters are connected, but borrowing weights are adaptively determined by covariate distances $d_{lj}$; $J_2$ and $J_3$ show strong similarity (thick lines), while $J_2$ and $J_4$ show weak similarity (dashed lines).}
\end{figure}

%% file: table/tableSWBB_sim_scn.tex
\begin{table}[tbp]
    \centering
    \fontsize{9}{12}\selectfont
    \caption{\label{SWBB:tab:sim:scn} Data generating mechanism for cluster-level mediation simulations.}
    \begin{tabular}{lp{10.5cm}} 
        \toprule
        Variable & \multicolumn{1}{c}{Data Generating Process} \\
        \midrule
        Random Effect & $\psi_{j} \sim N(0, 1)$ for $j = 1, \dots, J$ \\
        \cmidrule{1-2}
        Cluster Treatment & $Z_{j} \sim \text{Bern}(0.5)$ \\
        \cmidrule{1-2}
        Cluster Confounders & $V_{j1} \sim \text{Bin}(0.5)$; \quad $V_{j2} \sim \text{Bin}(0.4 + 0.2V_{j1})$; $V_{j3} \sim N(-1.5 + V_{j1} + 2V_{j2}, 2^2)$ \\
        \cmidrule{1-2}
        Individual Confounders & $C_{jiq} \sim \text{Bin}(\text{logit}^{-1}(\mathbb{V}_j \boldsymbol{\beta}_{jq}^c))$ for $q \in \{1,2,3\}$ \\
        & $\mathbf{C}_{jiq} \sim MVN(\mathbb{V}_j \boldsymbol{\beta}_{jq}^c, 0.7\mathbf{I}_3 + 0.3\mathbf{11}^\top)$ for $q \in \{4,5,6\}$ \\
        \cmidrule{1-2}
        \textbf{Scenario 1} ($\bbV \indep \bbC$) & Intercepts are $\{-1, 1, 0, -2, 2, 0\}$; all slopes set to 0 \\
        \cmidrule{1-2}
        \textbf{Scenario 2} (Weak $\bbV \to \bbC$) & $\boldsymbol{\beta}_{jq}^c = [\gamma_{jq}, 3, -2, -1]^\top$ \\
        \cmidrule{1-2}
        \textbf{Scenario 3} (Strong $\bbV \to \bbC$) & $\boldsymbol{\beta}_{jq}^c = [\gamma_{jq}, 6, -4, -2]^\top$ \\
        \cmidrule{1-2}
        Structural Models & $M_{j} \sim N(0.5 + Z_{j} + \bar{\mathbb{C}}_{j} \boldsymbol{\beta}^{mc} + \mathbb{V}_{j} \boldsymbol{\beta}^{mv}, 1^2)$ \\
        & $Y_{ji} \sim N(-20 + Z_{j} + M_{j} + \bar{\mathbb{C}}_{j} \boldsymbol{\beta}^{yc} + \mathbb{V}_{j} \boldsymbol{\beta}^{yv}, 2^2)$ \\
        \bottomrule
        \addlinespace
        \multicolumn{2}{l}{\textit{Note:} $\mathbb{V}_{j} = (1, V_{j1}, V_{j2}, V_{j3})^\top$. $q \in \{1, \dots, 6\}$ indexes the individual-level confounder dimension.}\\
        \multicolumn{2}{l}{For Scenarios 2--3, the intercept $\gamma_{jq}$ alternates sign based on base values: $\boldsymbol{\gamma}_{base} = \{ 0.5, 0.2, 0.3, 1.0, 0.4, 0.6 \}$.}\\
        \multicolumn{2}{l}{For $q \in \{1, 4\}$, $\gamma_{jq} = (-1)^j \gamma_{base,q}$. For $q \in \{2, 3, 5, 6\}$, $\gamma_{jq} = (-1)^{j+1} \gamma_{base,q}$.}
    \end{tabular}
\end{table}

%% file: table/tableSWBB_sim_results.tex
\begin{table}[tbp]
\fontsize{9}{10}\selectfont
\centering
\caption[Simulation results of causal effects for scenario 1, 2 and 3]{\label{SWBB:tab:sim:results} Simulation results for NIE, NDE, and ATE over 1,000 replications with a cluster-level continuous mediator. Results for the SWBB, HBB, and BB estimators under the GLM are displayed. The columns report the bias, RMSE of the posterior mean, and CP (empirical coverage probability), categorized by the strength of dependence between $\bbV$ and $\bbC$.}
\begin{tabular}{ll|ccc|ccc|ccc}
\toprule
 \multicolumn{2}{c}{} & \multicolumn{3}{c}{Scn 1 ($\mathbb{V} \perp \mathbb{C}$)} & \multicolumn{3}{c}{Scn 2 ($\mathbb{V} \not\perp \mathbb{C}$ - weak)} & \multicolumn{3}{c}{Scn 3 ($\mathbb{V} \not\perp \mathbb{C}$ - strong)} \\
 \cmidrule(lr){3-5} \cmidrule(lr){6-8} \cmidrule(lr){9-11}
 \multicolumn{1}{c}{$\chi$} & \multicolumn{1}{c}{$\zeta$} & Bias & RMSE & CP & Bias & RMSE & CP & Bias & RMSE & CP \\
\midrule
\multicolumn{2}{c}{NIE} & \multicolumn{3}{c}{True: 1.00} & \multicolumn{3}{c}{True: 1.00} & \multicolumn{3}{c}{True: 1.00} \\ \hline
BB & BB & 0.00 & 0.85 & 0.97 & -0.01 & 0.85 & 0.97 & -0.00 & 0.85 & 0.96 \\
HBB & HBB & 0.00 & 0.84 & 0.97 & -0.01 & 0.84 & 0.96 & -0.00 & 0.84 & 0.96 \\
0.01 & 0 & 0.00 & 0.85 & 0.97 & -0.01 & 0.85 & 0.97 & -0.00 & 0.85 & 0.96 \\
 & 0.5 & 0.00 & 0.85 & 0.97 & -0.01 & 0.85 & 0.97 & -0.00 & 0.85 & 0.96 \\
 & 1 & 0.00 & 0.84 & 0.97 & -0.01 & 0.84 & 0.96 & -0.00 & 0.84 & 0.96 \\
0.1 & 0 & 0.00 & 0.85 & 0.97 & -0.01 & 0.84 & 0.97 & -0.00 & 0.83 & 0.96 \\
 & 0.5 & 0.00 & 0.85 & 0.97 & -0.01 & 0.84 & 0.96 & -0.00 & 0.83 & 0.96 \\
 & 1 & 0.00 & 0.83 & 0.96 & -0.01 & 0.83 & 0.96 & -0.00 & 0.83 & 0.96 \\
1 & 0 & 0.00 & 0.81 & 0.96 & -0.01 & 0.81 & 0.96 & -0.00 & 0.81 & 0.95 \\
 & 0.5 & 0.00 & 0.81 & 0.96 & -0.01 & 0.81 & 0.96 & -0.00 & 0.81 & 0.95 \\
 & 1 & 0.00 & 0.81 & 0.96 & -0.01 & 0.81 & 0.95 & -0.00 & 0.81 & 0.95 \\
10 & 0 & 0.00 & 0.81 & 0.96 & -0.01 & 0.81 & 0.96 & -0.00 & 0.81 & 0.95 \\
 & 0.5 & 0.00 & 0.81 & 0.96 & -0.01 & 0.81 & 0.95 & -0.00 & 0.81 & 0.95 \\
 & 1 & 0.00 & 0.81 & 0.96 & -0.01 & 0.81 & 0.95 & -0.00 & 0.81 & 0.95 \\
100 & 0 & 0.00 & 0.81 & 0.96 & -0.01 & 0.81 & 0.96 & -0.00 & 0.81 & 0.95 \\
 & 0.5 & 0.00 & 0.81 & 0.96 & -0.01 & 0.81 & 0.95 & -0.00 & 0.81 & 0.95 \\
 & 1 & 0.00 & 0.81 & 0.96 & -0.01 & 0.81 & 0.95 & -0.00 & 0.81 & 0.95 \\
\midrule
\multicolumn{2}{c}{NDE} & \multicolumn{3}{c}{True: 1.00} & \multicolumn{3}{c}{True: 1.00} & \multicolumn{3}{c}{True: 1.00} \\ \hline
BB & BB & -0.01 & 0.75 & 0.95 & -0.02 & 0.75 & 0.95 & -0.01 & 0.75 & 0.95 \\
HBB & HBB & -0.01 & 0.75 & 0.95 & -0.02 & 0.75 & 0.95 & -0.01 & 0.75 & 0.95 \\
0.01 & 0 & -0.01 & 0.75 & 0.95 & -0.02 & 0.75 & 0.95 & -0.01 & 0.75 & 0.95 \\
 & 0.5 & -0.01 & 0.75 & 0.95 & -0.02 & 0.75 & 0.95 & -0.01 & 0.75 & 0.95 \\
 & 1 & -0.01 & 0.75 & 0.95 & -0.02 & 0.75 & 0.95 & -0.01 & 0.75 & 0.95 \\
0.1 & 0 & -0.01 & 0.75 & 0.95 & -0.02 & 0.75 & 0.95 & -0.01 & 0.75 & 0.95 \\
 & 0.5 & -0.01 & 0.75 & 0.95 & -0.02 & 0.75 & 0.95 & -0.01 & 0.75 & 0.95 \\
 & 1 & -0.01 & 0.75 & 0.95 & -0.02 & 0.75 & 0.95 & -0.01 & 0.75 & 0.95 \\
1 & 0 & -0.01 & 0.74 & 0.95 & -0.02 & 0.74 & 0.94 & -0.01 & 0.75 & 0.94 \\
 & 0.5 & -0.01 & 0.74 & 0.95 & -0.02 & 0.74 & 0.95 & -0.01 & 0.75 & 0.95 \\
 & 1 & -0.01 & 0.74 & 0.95 & -0.02 & 0.74 & 0.94 & -0.01 & 0.75 & 0.95 \\
10 & 0 & -0.01 & 0.74 & 0.95 & -0.02 & 0.74 & 0.94 & -0.01 & 0.75 & 0.95 \\
 & 0.5 & -0.01 & 0.74 & 0.95 & -0.02 & 0.74 & 0.94 & -0.01 & 0.75 & 0.94 \\
 & 1 & -0.01 & 0.74 & 0.95 & -0.02 & 0.74 & 0.94 & -0.01 & 0.75 & 0.94 \\
100 & 0 & -0.01 & 0.74 & 0.95 & -0.02 & 0.74 & 0.94 & -0.01 & 0.75 & 0.95 \\
 & 0.5 & -0.01 & 0.74 & 0.95 & -0.02 & 0.74 & 0.95 & -0.01 & 0.75 & 0.95 \\
 & 1 & -0.01 & 0.74 & 0.95 & -0.02 & 0.74 & 0.94 & -0.01 & 0.75 & 0.95 \\
\midrule
\multicolumn{2}{c}{ATE} & \multicolumn{3}{c}{True: 2.00} & \multicolumn{3}{c}{True: 2.00} & \multicolumn{3}{c}{True: 2.00} \\ \hline
BB & BB & -0.01 & 0.91 & 0.97 & -0.03 & 0.91 & 0.97 & -0.01 & 0.91 & 0.97 \\
HBB & HBB & -0.01 & 0.90 & 0.97 & -0.03 & 0.90 & 0.97 & -0.01 & 0.90 & 0.96 \\
0.01 & 0 & -0.01 & 0.91 & 0.97 & -0.03 & 0.91 & 0.97 & -0.01 & 0.91 & 0.97 \\
 & 0.5 & -0.01 & 0.91 & 0.97 & -0.03 & 0.91 & 0.97 & -0.01 & 0.91 & 0.97 \\
 & 1 & -0.01 & 0.90 & 0.97 & -0.03 & 0.90 & 0.97 & -0.01 & 0.90 & 0.96 \\
0.1 & 0 & -0.01 & 0.91 & 0.97 & -0.03 & 0.90 & 0.97 & -0.01 & 0.89 & 0.96 \\
 & 0.5 & -0.01 & 0.91 & 0.97 & -0.03 & 0.90 & 0.97 & -0.01 & 0.89 & 0.97 \\
 & 1 & -0.01 & 0.89 & 0.97 & -0.03 & 0.89 & 0.96 & -0.01 & 0.89 & 0.96 \\
1 & 0 & -0.01 & 0.87 & 0.96 & -0.03 & 0.88 & 0.96 & -0.01 & 0.88 & 0.95 \\
 & 0.5 & -0.01 & 0.87 & 0.96 & -0.03 & 0.88 & 0.96 & -0.01 & 0.88 & 0.96 \\
 & 1 & -0.01 & 0.88 & 0.96 & -0.03 & 0.88 & 0.96 & -0.01 & 0.88 & 0.96 \\
10 & 0 & -0.01 & 0.87 & 0.96 & -0.03 & 0.87 & 0.96 & -0.01 & 0.88 & 0.96 \\
 & 0.5 & -0.01 & 0.87 & 0.96 & -0.03 & 0.87 & 0.96 & -0.01 & 0.88 & 0.96 \\
 & 1 & -0.01 & 0.87 & 0.96 & -0.03 & 0.87 & 0.96 & -0.01 & 0.88 & 0.96 \\
100 & 0 & -0.01 & 0.87 & 0.96 & -0.03 & 0.87 & 0.96 & -0.01 & 0.88 & 0.95 \\
 & 0.5 & -0.01 & 0.87 & 0.96 & -0.03 & 0.87 & 0.96 & -0.01 & 0.88 & 0.95 \\
 & 1 & -0.01 & 0.87 & 0.96 & -0.03 & 0.87 & 0.96 & -0.01 & 0.88 & 0.95 \\
\midrule
\bottomrule
\end{tabular}
\end{table}

%% file: table/tableSWBB_BIGPIC_Z1vsZ3_m12.tex
\begin{table}[tbp]
\centering
\caption{\label{SWBB:tab:Z1vsZ3_main} Causal Mediation Analysis Results for the Additive Effect of Group Medical Visits (GMV) within Microfinance (MF) Networks. The table presents posterior estimates for the Natural Indirect Effect (NIE), Natural Direct Effect (NDE), and Average Total Effect (ATE) comparing the usual care plus microfinance arm ($Z=1$) against the integrated group medical visits and microfinance arm ($Z=3$: GMV-MF). Results are evaluated across four sociometric network mediators measured at 12 months: Density, Transitivity, Cohesion, and Average Path Length (APL). We compare three Bayesian Bootstrap specifications: standard (BB), hierarchical (HBB), and the proposed two-stage similarity-based (SWBB). Metrics include the posterior mean (Est), posterior standard deviation (SD), 95\% credible intervals (2.5\%, 97.5\ and the credible interval length (CIl). Conditional effects for the SWBB are provided for community-level microfinance penetration—(1) Light and (2) Heavy GISE—and individual-level diabetes status—(3) Non-diabetic and (4) Diabetic.}
\resizebox{\ifdim\width>\linewidth\linewidth\else\width\fi}{!}{
\fontsize{12}{12}\selectfont
\renewcommand{\arraystretch}{1.25}
\begin{tabular}[t]{lcrrrrrrrrrrrrrrr}
\toprule
\multicolumn{2}{c}{ } & \multicolumn{5}{c}{Natural Indirect Effect} & \multicolumn{5}{c}{Natural Direct Effect} & \multicolumn{5}{c}{Average Total Effect} \\
\cmidrule(l{3pt}r{3pt}){3-7} \cmidrule(l{3pt}r{3pt}){8-12} \cmidrule(l{3pt}r{3pt}){13-17}
Method & Cond. & Est & SD & 2.5\% & 97.5\% & CIl & Est & SD & 2.5\% & 97.5\% & CIl & Est & SD & 2.5\% & 97.5\% & CIl\\
\midrule
\addlinespace[0.3em]
\multicolumn{17}{l}{\textbf{Mediator: Density}}\\
\hline
\hspace{1em}\textbf{BB} & Pop & -0.01 & 1.17 & -2.35 & 2.17 & 4.52 & -0.09 & 1.35 & -2.40 & 2.57 & 4.97 & -0.10 & 1.37 & -2.84 & 2.63 & 5.48\\
\hspace{1em}\textbf{HBB} & Pop & 0.20 & 1.23 & -1.99 & 2.56 & 4.55 & -0.18 & 1.28 & -2.59 & 2.41 & 5.00 & 0.01 & 1.38 & -2.78 & 2.73 & 5.51\\
\hspace{1em}\textbf{SWBB} & Pop & -0.18 & 0.93 & -1.92 & 1.49 & 3.42 & 0.05 & 1.21 & -2.22 & 2.55 & 4.78 & -0.13 & 1.19 & -2.49 & 1.80 & 4.29\\
\cmidrule{1-17}
\hspace{1em}\textbf{SWBB} & (1) & -0.17 & 1.44 & -3.01 & 2.54 & 5.55 & 0.10 & 1.58 & -2.77 & 3.42 & 6.20 & -0.07 & 1.53 & -3.43 & 2.67 & 6.10\\
\hspace{1em}\textbf{SWBB} & (2) & 0.00 & 1.41 & -2.93 & 2.62 & 5.55 & 0.04 & 1.78 & -3.14 & 3.32 & 6.46 & 0.05 & 1.78 & -3.04 & 3.30 & 6.33\\
\hspace{1em}\textbf{SWBB} & (3) & 0.07 & 1.16 & -2.34 & 1.96 & 4.30 & -0.00 & 1.32 & -2.17 & 2.90 & 5.07 & 0.07 & 1.23 & -2.50 & 2.19 & 4.69\\
\hspace{1em}\textbf{SWBB} & (4) & 0.09 & 2.39 & -4.63 & 4.23 & 8.86 & 0.12 & 2.52 & -4.93 & 5.52 & 10.45 & 0.20 & 2.59 & -4.83 & 5.01 & 9.84\\
\addlinespace[0.3em]
\hline
\multicolumn{17}{l}{\textbf{Mediator: Transitivity}}\\
\hline
\hspace{1em}\textbf{BB} & Pop & -0.01 & 1.18 & -2.34 & 2.15 & 4.49 & -0.09 & 1.36 & -2.42 & 2.60 & 5.02 & -0.10 & 1.37 & -2.79 & 2.62 & 5.41\\
\hspace{1em}\textbf{HBB} & Pop & 0.20 & 1.23 & -1.99 & 2.59 & 4.59 & -0.18 & 1.28 & -2.58 & 2.42 & 5.01 & 0.01 & 1.38 & -2.76 & 2.73 & 5.49\\
\hspace{1em}\textbf{SWBB} & Pop & -0.18 & 0.93 & -1.93 & 1.50 & 3.43 & 0.05 & 1.21 & -2.19 & 2.57 & 4.76 & -0.13 & 1.20 & -2.52 & 1.81 & 4.33\\
\cmidrule{1-17}
\hspace{1em}\textbf{SWBB} & (1) & -0.17 & 1.44 & -3.01 & 2.52 & 5.53 & 0.10 & 1.58 & -2.79 & 3.41 & 6.20 & -0.07 & 1.53 & -3.46 & 2.69 & 6.15\\
\hspace{1em}\textbf{SWBB} & (2) & -0.00 & 1.41 & -2.93 & 2.62 & 5.55 & 0.04 & 1.78 & -3.14 & 3.31 & 6.45 & 0.04 & 1.78 & -3.04 & 3.33 & 6.36\\
\hspace{1em}\textbf{SWBB} & (3) & 0.07 & 1.16 & -2.34 & 1.96 & 4.30 & -0.00 & 1.32 & -2.18 & 2.92 & 5.09 & 0.07 & 1.24 & -2.51 & 2.20 & 4.71\\
\hspace{1em}\textbf{SWBB} & (4) & 0.09 & 2.39 & -4.62 & 4.24 & 8.86 & 0.12 & 2.52 & -4.93 & 5.53 & 10.46 & 0.20 & 2.59 & -4.81 & 5.00 & 9.81\\
\addlinespace[0.3em]
\hline
\multicolumn{17}{l}{\textbf{Mediator: Cohesion}}\\
\hline
\hspace{1em}\textbf{BB} & Pop & -0.01 & 1.18 & -2.35 & 2.12 & 4.47 & -0.09 & 1.36 & -2.41 & 2.69 & 5.10 & -0.10 & 1.37 & -2.72 & 2.58 & 5.30\\
\hspace{1em}\textbf{HBB} & Pop & 0.20 & 1.24 & -2.00 & 2.64 & 4.64 & -0.18 & 1.29 & -2.57 & 2.36 & 4.93 & 0.02 & 1.38 & -2.73 & 2.71 & 5.44\\
\hspace{1em}\textbf{SWBB} & Pop & -0.18 & 0.93 & -1.95 & 1.49 & 3.44 & 0.05 & 1.22 & -2.16 & 2.63 & 4.79 & -0.13 & 1.20 & -2.62 & 1.82 & 4.44\\
\cmidrule{1-17}
\hspace{1em}\textbf{SWBB} & (1) & -0.17 & 1.44 & -2.97 & 2.41 & 5.38 & 0.10 & 1.59 & -2.84 & 3.34 & 6.18 & -0.07 & 1.55 & -3.46 & 2.73 & 6.20\\
\hspace{1em}\textbf{SWBB} & (2) & -0.01 & 1.42 & -2.96 & 2.61 & 5.57 & 0.04 & 1.78 & -3.13 & 3.33 & 6.46 & 0.04 & 1.78 & -3.09 & 3.35 & 6.43\\
\hspace{1em}\textbf{SWBB} & (3) & 0.08 & 1.17 & -2.34 & 1.96 & 4.30 & -0.00 & 1.32 & -2.19 & 2.94 & 5.12 & 0.08 & 1.25 & -2.46 & 2.20 & 4.66\\
\hspace{1em}\textbf{SWBB} & (4) & 0.09 & 2.39 & -4.61 & 4.26 & 8.86 & 0.12 & 2.52 & -4.94 & 5.56 & 10.51 & 0.20 & 2.60 & -4.83 & 5.04 & 9.87\\
\addlinespace[0.3em]
\hline
\multicolumn{17}{l}{\textbf{Mediator: APL}}\\
\hline
\hspace{1em}\textbf{BB} & Pop & -0.00 & 1.17 & -2.34 & 2.22 & 4.56 & -0.09 & 1.35 & -2.39 & 2.50 & 4.89 & -0.09 & 1.37 & -2.85 & 2.65 & 5.50\\
\hspace{1em}\textbf{HBB} & Pop & 0.20 & 1.23 & -1.97 & 2.53 & 4.50 & -0.18 & 1.28 & -2.62 & 2.36 & 4.98 & 0.02 & 1.38 & -2.81 & 2.80 & 5.61\\
\hspace{1em}\textbf{SWBB} & Pop & -0.18 & 0.93 & -1.97 & 1.51 & 3.47 & 0.05 & 1.21 & -2.29 & 2.52 & 4.80 & -0.13 & 1.19 & -2.48 & 1.82 & 4.30\\
\cmidrule{1-17}
\hspace{1em}\textbf{SWBB} & (1) & -0.17 & 1.44 & -2.98 & 2.60 & 5.58 & 0.10 & 1.58 & -2.77 & 3.45 & 6.22 & -0.07 & 1.53 & -3.39 & 2.62 & 6.01\\
\hspace{1em}\textbf{SWBB} & (2) & 0.00 & 1.41 & -2.95 & 2.62 & 5.57 & 0.04 & 1.78 & -3.16 & 3.35 & 6.51 & 0.05 & 1.78 & -3.04 & 3.30 & 6.34\\
\hspace{1em}\textbf{SWBB} & (3) & 0.07 & 1.16 & -2.33 & 1.97 & 4.29 & -0.00 & 1.32 & -2.20 & 2.86 & 5.06 & 0.07 & 1.23 & -2.47 & 2.12 & 4.59\\
\hspace{1em}\textbf{SWBB} & (4) & 0.09 & 2.39 & -4.61 & 4.27 & 8.88 & 0.11 & 2.52 & -4.89 & 5.49 & 10.38 & 0.20 & 2.59 & -4.82 & 5.05 & 9.87\\
\addlinespace[0.3em]
\hline
\multicolumn{17}{l}{\rule{0pt}{1em}\textit{Note: Pop: Population Average. Conditional Effects: (1) GISE = light, (2) GISE = heavy, (3) DM = 0 (No), (4) DM = 1 (Yes).}} \\
\bottomrule
\end{tabular}}
\end{table}

%% file: table/tableSWBB_BIGPIC_Z2vsZ3_m12.tex
\begin{table}[tbp]
\centering
\caption{\label{SWBB:tab:Z2vsZ3_main} Causal Mediation Analysis Results for the Effect of Microfinance (MF) within Group Medical Visit (GMV) Networks. The table presents posterior estimates for the Natural Indirect Effect (NIE), Natural Direct Effect (NDE), and Average Total Effect (ATE) comparing the group medical visits arm ($Z=2$) against the integrated group medical visits and microfinance arm ($Z=3$: GMV-MF). This comparison evaluates the impact of adding microfinance to an existing medical visit framework. Results are evaluated across four sociometric network mediators measured at 12 months: Density, Transitivity, Cohesion, and Average Path Length (APL). We compare three Bayesian Bootstrap specifications: standard (BB), hierarchical (HBB), and the proposed two-stage similarity-based (SWBB). Metrics include the posterior mean (Est), posterior standard deviation (SD), 95\% credible intervals (2.5\%, 97.5\%), and the credible interval length (CIl). Conditional effects for the SWBB are provided for community-level microfinance penetration—(1) Light and (2) Heavy GISE—and individual-level diabetes status—(3) Non-diabetic and (4) Diabetic.}
\resizebox{\ifdim\width>\linewidth\linewidth\else\width\fi}{!}{
\fontsize{12}{12}\selectfont
\renewcommand{\arraystretch}{1.25}
\begin{tabular}[t]{lcrrrrrrrrrrrrrrr}
\toprule
\multicolumn{2}{c}{ } & \multicolumn{5}{c}{Natural Indirect Effect} & \multicolumn{5}{c}{Natural Direct Effect} & \multicolumn{5}{c}{Average Total Effect} \\
\cmidrule(l{3pt}r{3pt}){3-7} \cmidrule(l{3pt}r{3pt}){8-12} \cmidrule(l{3pt}r{3pt}){13-17}
Method & Cond. & Est & SD & 2.5\% & 97.5\% & CIl & Est & SD & 2.5\% & 97.5\% & CIl & Est & SD & 2.5\% & 97.5\% & CIl\\
\midrule
\addlinespace[0.3em]
\multicolumn{17}{l}{\textbf{Mediator: Density}}\\
\hline
\hspace{1em}\textbf{BB} & Pop & 0.12 & 1.41 & -2.53 & 3.02 & 5.55 & -0.00 & 1.51 & -2.87 & 3.07 & 5.94 & 0.11 & 1.51 & -2.77 & 3.25 & 6.02\\
\hspace{1em}\textbf{HBB} & Pop & -0.09 & 1.38 & -2.69 & 2.59 & 5.29 & 0.01 & 1.43 & -2.81 & 2.66 & 5.47 & -0.08 & 1.47 & -3.04 & 2.61 & 5.64\\
\hspace{1em}\textbf{SWBB} & Pop & 0.02 & 1.07 & -1.92 & 1.93 & 3.86 & -0.06 & 1.23 & -2.33 & 2.18 & 4.52 & -0.04 & 1.16 & -2.55 & 2.11 & 4.66\\
\cmidrule{1-17}
\hspace{1em}\textbf{SWBB} & (1) & 0.13 & 1.37 & -2.69 & 2.71 & 5.39 & -0.10 & 1.40 & -2.85 & 2.54 & 5.39 & 0.02 & 1.38 & -2.73 & 2.38 & 5.11\\
\hspace{1em}\textbf{SWBB} & (2) & -0.30 & 1.96 & -3.95 & 3.25 & 7.20 & 0.23 & 2.06 & -3.88 & 3.75 & 7.64 & -0.08 & 2.02 & -3.77 & 3.63 & 7.40\\
\hspace{1em}\textbf{SWBB} & (3) & -0.18 & 1.21 & -2.71 & 2.23 & 4.95 & 0.15 & 1.41 & -2.24 & 3.01 & 5.25 & -0.03 & 1.30 & -2.83 & 2.74 & 5.58\\
\hspace{1em}\textbf{SWBB} & (4) & -0.16 & 2.31 & -4.53 & 4.61 & 9.15 & -0.02 & 2.37 & -4.51 & 4.74 & 9.25 & -0.18 & 2.49 & -4.99 & 4.86 & 9.85\\
\addlinespace[0.3em]
\hline
\multicolumn{17}{l}{\textbf{Mediator: Transitivity}}\\
\hline
\hspace{1em}\textbf{BB} & Pop & 0.12 & 1.41 & -2.54 & 3.04 & 5.58 & -0.00 & 1.51 & -2.87 & 3.07 & 5.94 & 0.11 & 1.51 & -2.75 & 3.22 & 5.97\\
\hspace{1em}\textbf{HBB} & Pop & -0.09 & 1.38 & -2.69 & 2.59 & 5.28 & 0.01 & 1.43 & -2.81 & 2.67 & 5.47 & -0.07 & 1.47 & -3.02 & 2.61 & 5.63\\
\hspace{1em}\textbf{SWBB} & Pop & 0.02 & 1.06 & -1.91 & 1.91 & 3.82 & -0.06 & 1.23 & -2.33 & 2.18 & 4.52 & -0.04 & 1.16 & -2.55 & 2.11 & 4.65\\
\cmidrule{1-17}
\hspace{1em}\textbf{SWBB} & (1) & 0.13 & 1.37 & -2.70 & 2.71 & 5.41 & -0.10 & 1.40 & -2.85 & 2.54 & 5.39 & 0.02 & 1.38 & -2.74 & 2.38 & 5.12\\
\hspace{1em}\textbf{SWBB} & (2) & -0.30 & 1.96 & -3.97 & 3.26 & 7.23 & 0.23 & 2.06 & -3.88 & 3.76 & 7.64 & -0.08 & 2.02 & -3.75 & 3.65 & 7.40\\
\hspace{1em}\textbf{SWBB} & (3) & -0.18 & 1.21 & -2.71 & 2.21 & 4.93 & 0.15 & 1.41 & -2.24 & 3.01 & 5.25 & -0.03 & 1.30 & -2.85 & 2.74 & 5.60\\
\hspace{1em}\textbf{SWBB} & (4) & -0.16 & 2.31 & -4.52 & 4.61 & 9.13 & -0.02 & 2.37 & -4.51 & 4.74 & 9.25 & -0.18 & 2.49 & -4.99 & 4.86 & 9.84\\
\addlinespace[0.3em]
\hline
\multicolumn{17}{l}{\textbf{Mediator: Cohesion}}\\
\hline
\hspace{1em}\textbf{BB} & Pop & 0.12 & 1.41 & -2.54 & 3.08 & 5.62 & -0.00 & 1.51 & -2.88 & 3.07 & 5.94 & 0.11 & 1.51 & -2.73 & 3.19 & 5.91\\
\hspace{1em}\textbf{HBB} & Pop & -0.08 & 1.38 & -2.68 & 2.58 & 5.26 & 0.01 & 1.43 & -2.79 & 2.68 & 5.47 & -0.07 & 1.46 & -3.04 & 2.63 & 5.68\\
\hspace{1em}\textbf{SWBB} & Pop & 0.02 & 1.06 & -1.92 & 1.91 & 3.83 & -0.06 & 1.23 & -2.33 & 2.19 & 4.51 & -0.04 & 1.16 & -2.53 & 2.10 & 4.62\\
\cmidrule{1-17}
\hspace{1em}\textbf{SWBB} & (1) & 0.13 & 1.37 & -2.70 & 2.72 & 5.42 & -0.10 & 1.40 & -2.85 & 2.54 & 5.39 & 0.02 & 1.38 & -2.74 & 2.39 & 5.13\\
\hspace{1em}\textbf{SWBB} & (2) & -0.30 & 1.96 & -3.97 & 3.28 & 7.25 & 0.23 & 2.06 & -3.88 & 3.77 & 7.64 & -0.08 & 2.02 & -3.73 & 3.65 & 7.38\\
\hspace{1em}\textbf{SWBB} & (3) & -0.18 & 1.21 & -2.71 & 2.19 & 4.90 & 0.15 & 1.41 & -2.25 & 3.01 & 5.25 & -0.03 & 1.31 & -2.88 & 2.74 & 5.62\\
\hspace{1em}\textbf{SWBB} & (4) & -0.16 & 2.31 & -4.52 & 4.61 & 9.13 & -0.02 & 2.37 & -4.50 & 4.74 & 9.25 & -0.18 & 2.50 & -4.98 & 4.85 & 9.83\\
\addlinespace[0.3em]
\hline
\multicolumn{17}{l}{\textbf{Mediator: APL}}\\
\hline
\hspace{1em}\textbf{BB} & Pop & 0.12 & 1.42 & -2.54 & 2.99 & 5.53 & -0.00 & 1.51 & -2.86 & 3.07 & 5.93 & 0.12 & 1.51 & -2.79 & 3.30 & 6.09\\
\hspace{1em}\textbf{HBB} & Pop & -0.09 & 1.38 & -2.70 & 2.57 & 5.27 & 0.01 & 1.43 & -2.82 & 2.66 & 5.48 & -0.08 & 1.47 & -3.11 & 2.61 & 5.72\\
\hspace{1em}\textbf{SWBB} & Pop & 0.02 & 1.07 & -1.93 & 1.97 & 3.89 & -0.06 & 1.23 & -2.34 & 2.18 & 4.52 & -0.04 & 1.16 & -2.56 & 2.12 & 4.68\\
\cmidrule{1-17}
\hspace{1em}\textbf{SWBB} & (1) & 0.13 & 1.37 & -2.70 & 2.71 & 5.40 & -0.10 & 1.40 & -2.85 & 2.54 & 5.39 & 0.02 & 1.38 & -2.74 & 2.38 & 5.12\\
\hspace{1em}\textbf{SWBB} & (2) & -0.30 & 1.96 & -3.98 & 3.26 & 7.24 & 0.23 & 2.06 & -3.89 & 3.75 & 7.63 & -0.08 & 2.02 & -3.80 & 3.57 & 7.37\\
\hspace{1em}\textbf{SWBB} & (3) & -0.18 & 1.21 & -2.73 & 2.23 & 4.96 & 0.15 & 1.41 & -2.23 & 3.01 & 5.24 & -0.03 & 1.30 & -2.79 & 2.74 & 5.52\\
\hspace{1em}\textbf{SWBB} & (4) & -0.16 & 2.30 & -4.52 & 4.63 & 9.15 & -0.02 & 2.37 & -4.50 & 4.74 & 9.25 & -0.19 & 2.49 & -5.00 & 4.86 & 9.87\\
\addlinespace[0.3em]
\hline
\multicolumn{17}{l}{\rule{0pt}{1em}\textit{Note: Pop: Population Average. Conditional Effects: (1) GISE = light, (2) GISE = heavy, (3) DM = 0 (No), (4) DM = 1 (Yes).}} \\
\bottomrule
\end{tabular}}
\end{table}